\documentclass[a4paper,12pt]{article}
\usepackage{amsfonts}
\usepackage{amsmath}
\usepackage{amssymb}
\usepackage{graphicx}
\usepackage[utf8]{inputenc}
\usepackage[english]{babel}
\usepackage{ragged2e}
\usepackage{indentfirst}
\usepackage{lipsum}
\DeclareRobustCommand{\rchi}{{\mathpalette\irchi\relax}}
\newcommand{\irchi}[2]{\raisebox{\depth}{$#1k$}}
\usepackage{fullpage}
\usepackage{lscape}
\usepackage{wrapfig}
\usepackage{tensor}
\usepackage{makeidx}
\usepackage{url}
\usepackage[margin=0.8in]{geometry}
\usepackage{pdfpages}
\usepackage{xcolor}
\usepackage{authblk}
\usepackage[]{hyperref}
\usepackage{mcite}
\hypersetup{
	colorlinks=true,
	linkcolor={blue},
	citecolor={cyan!80!black},
	filecolor={red!60!green},      
	urlcolor={magenta!60!black},
}
\newcommand*{\bfrac}[2]{\genfrac{\lbrace}{\rbrace}{0pt}{}{#1}{#2}}
\title{Non-relativistic limit of
	Einstein-Cartan-Dirac equations}

\author[1]{Swanand Khanapurkar}
\author[2]{Arnab Pradhan}
\author[3]{Vedant Dhruv}
\author[4]{Tejinder P. Singh}
\affil[1]{\textit{Indian Institute of Science Education and Research (IISER), Pune 411008, India.}}
\affil[2]{\textit{Indian Institute of Technology Madras (IITM), Chennai 600036, India.}}
\affil[3]{\textit{National Institute of Technology Karnataka (NITK), Surathkal 575025, India}}
\affil[4]{\textit{Tata Institute of Fundamental Research (TIFR), Mumbai 400005, India.}}

\date{}

\begin{document}

\maketitle

\texttt{E-mail: $^1$\url{swanand.khanapurkar@students.iiserpune.ac.in}, $^2$\url{arnab.science@hotmail.com}, $^3$\url{vedantdhruv96@gmail.com},$^4$\url{tpsingh@tifr.res.in}}
\bigskip
\begin{abstract}
\noindent We derive the Schr\"{o}dinger-Newton equation as the non-relativistic limit of the Einstein-Dirac equations. Our
analysis relaxes the assumption of spherical symmetry, made in an earlier work in the literature, while deriving this limit. Since
the spin of the Dirac field couples naturally to torsion, we generalize our analysis to the Einstein-Cartan-Dirac (ECD) equations, again recovering the Schr\"{o}dinger-Newton equation.

\end{abstract}
\def\baselinestretch{1}\selectfont
\noindent\rule{17 cm}{0.6pt}
\tableofcontents
\noindent\rule{17 cm}{0.6pt}

\section{Introduction}\label{intro}
The Schr\"{o}dinger-Newton [SN] equation has been proposed in the literature as a model for investigating the effects of self-gravity on the motion of a non-relativistic quantum particle \cite{SN_eqn_originalidea,SN_diosi,SN_penrose_1,SN_penrose_2} (specifically as a model for gravitational localization of macroscopic objects).
It is a nonlinear modification to the Schr\"odinger equation with a Newtonian gravitational potential $\phi$:
\begin{equation}
	i\hbar \frac{\partial \psi(\textbf{r},t)}{\partial t} = - \frac{\hbar^2}{2m} \nabla^{2}\psi(\textbf{r},t) + m\phi~\psi(\textbf{r},t), \label{3}
\end{equation}
where the self-gravitating potential $\phi$ is assumed to be classical and obeys the semi-classical Poisson equation
\begin{equation}\label{2}
	\nabla^{2} \phi = 4\pi G m \vert \psi \vert ^{2}.
\end{equation}
The coupled system of the above two equations in integro-differential form is given by
\begin{equation}
	i\hbar \frac{\partial \psi(\textbf{r},t)}{\partial t} = - \frac{\hbar^2}{2m} \nabla^{2}\psi(\textbf{r},t) - Gm^{2} \int \frac{\vert\psi(\textbf{r}',t)\vert^{2}}{\vert \textbf{r} - \textbf{r}'\vert} d^{3}r' \psi(\textbf{r},t), \label{4}
\end{equation}
and is known as the Schr\"{o}dinger-Newton [SN] equation.
There are two broad (complementary) viewpoints under which the SN equation has been dealt with in the literature, amongst others. In one of them, it is considered as a hypothesis and the ways to falsify it are studied through theoretical and (or) experimental considerations, e.g. the localization of wave-packets for macroscopic objects \cite{giulini_SN_gravity_despersion}, with a gravitationally induced inhibition of quantum dispersion. The second approach focuses on whether the SN equation can be understood as a consequence of the known principles of physics. It is viewed as a model for self-interaction of matter waves. Notable work in this context \cite{guilini_grosardt} shows that the SN equation is the non-relativistic limit of the Einstein-Klein-Gordon system and the Einstein-Dirac system for a spherically symmetric space-time. Our present paper follows the second approach. We relax the assumption of a spherically symmetric space-time made in \cite{guilini_grosardt} and obtain the SN equation as the non-relativistic limit of the Einstein-Dirac equations. Since the spin of the Dirac field couples naturally to torsion, we also study the Einstein-Cartan-Dirac equations, and obtain it's non-relativistic limit. These equations are a special case of the Einstein-Cartan-Sciama-Kibble theory \cite{Cartan_1922, trautman_ECT, Hehl1971, hehl_RMP, sciama_physica_structure_of_GR, kibble_GR, gasperini, sciama_GR}, which we will henceforth refer to as the Einstein-Cartan theory. 

The plan of the paper is as follows. In Section 2 we describe the Einstein-Cartan-Dirac equations. Section 3 is the central part of the paper - the non-relativistic limit of the Einstein-Dirac equations is derived here. We first describe the ansatz  for the Dirac state and for the metric, which is used to derive the non-relativistic limit. We then describe in detail the non-relativistic expansion for the Dirac equation, and for the energy-momentum tensor. It is then shown that the non-relativistic limit of the Eistein-Dirac equations is the Schr\"{o}dinger-Newton equation, as expected. In Section 4, the non-relativistic limit of the Einstein-Cartan-Dirac equations - which include the torsion of the Dirac field - is derived. It is shown that torsion does not contribute in the non-relativistic limit, and once again we obtain the Schr\"{o}dinger-Newton equation. Conclusions are presented in the next section, while the detailed Appendix gives calculations of the geometric variables such as metric, connection and curvature, as well as the energy-momentum tensor, for the ansatz used in this paper.

The present paper is part of a series of our works \cite{Anushrut1, Anushrut2, tps1, newlength1, newlength2, swanand1, swanand2} which investigate the role of torsion in microscopic physics, and the motivation for including torsion in the Einstein-Dirac equations.
The fundamental motivation comes from noting that given a relativistic point mass $m$, Einstein equations as well as the Dirac equation both claim to hold for it, irrespective of the numerical value of the mass. This is because there is no mass scale in either of the system of equations, but of course both cannot hold for all masses. Only from experiments we know that Einstein equations hold for macroscopic masses, and Dirac equation for small masses. But how large is large, and how small is small? There has to be an underlying dynamics with an inbuilt mass scale, to which the Dirac equation and Einstein equation are small mass and large mass approximations, respectively. The search for this underlying dynamics is aided by the fact that general relativity has Schwarzschild radius as a fundamental length (depending linearly on mass) and Dirac equation has Compton wavelength as fundamental length (depending inversely on mass). This strongly suggests that the underlying theory should have one unified length, and also that it should include torsion, which dominates over curvature for small masses, because in this domain spin dominates mass. We have developed such curvature-torsion models, and investigated what physical role torsion might play in the modified Dirac equation. It is in this spirit that in this paper  we are studying the non-relativistic limit of the Eistein-Cartan-Dirac equations, to look for signatures of torsion. 

\section{Preliminaries: The Einstein-Cartan-Dirac equations}\label{EC_dirac_coupling}
The antisymmetric part of the affine connection,
\begin{equation}\label{eq:torsion tensor}
Q_{\alpha\beta}^{\:\:\:\:\:\:\mu} = \Gamma_{[\alpha\beta]}^{\:\:\:\:\:\:\:\mu} = \frac{1}{2}(\Gamma_{\alpha\beta}^{\:\:\:\:\:\:\mu} - \Gamma_{\beta\alpha}^{\:\:\:\:\:\:\mu}),
\end{equation} 
is called torsion. The affine connection is related to the Christoffel symbols by
\begin{align}\label{eq:affine connection in terms of contorsion}
\Gamma_{\alpha\beta}^{\:\:\:\:\:\:\:\mu} = \bfrac{\mu}{\alpha\beta}- K_{\alpha\beta}^{\:\:\:\:\:\:\:\mu},
\end{align} 
where $K_{\alpha\beta}^{\:\:\:\:\:\:\:\mu}$ is the contorsion tensor, and is given by $K_{\alpha\beta}^{\:\:\:\:\:\:\:\mu} = -Q_{\alpha\beta}^{\:\:\:\:\:\:\:\mu} - Q^{\mu}_{\:\:\:\alpha\beta} + Q_{\beta \:\:\:\alpha}^{\:\:\:\mu}$.

For a matter field $\psi$, which is minimally coupled to gravity and torsion, the action is given by \cite{hehl_RMP}
\begin{equation} \label{generic S for EC-matter coupling}
S = \int d^{4}x \sqrt{-g} \Big[\mathcal{L}_{m} (\psi, \nabla\psi, g) - \frac{1}{2\rchi} R(g, \partial g, Q)\Big],
\end{equation}
where $\rchi = {8\pi G}/{c^4}$. The first and the second term on the right hand side correspond to contribution from matter and gravity respectively. Varying the action with respect to $\psi$ (matter field), $g_{\mu\nu}$ (metric) and $K_{\alpha\beta\mu}$ (contorsion), the following field equations are obtained:
\begin{align}
\frac{\delta(\sqrt{-g}\mathcal{L}_m)}{\delta \psi} &= 0, \label{eq:generic EOM of matter}\\
\frac{\delta(\sqrt{-g}R)}{\delta g^{\mu\nu}} &= 2\rchi \frac{\delta(\sqrt{-g}\mathcal{L}_{m})}{\delta g^{\mu\nu}}        \ \text{and} \label{eq:generic EOM of metric-mass}\\
\frac{\delta(\sqrt{-g}R)}{\delta K_{\alpha\beta\mu}} &= 2\rchi \frac{\delta(\sqrt{-g}\mathcal{L}_{m})}{\delta K_{\alpha\beta\mu}}. \label{eq:generic EOM of torsion-spin}
\end{align}
Eq. (\ref{eq:generic EOM of matter}) yields the matter field equation on a space-time with torsion.
The right hand side of Eq. (\ref{eq:generic EOM of metric-mass}) is related to the metric energy-momentum tensor $T_{\mu\nu}$, while the right hand side  of Eq. (\ref{eq:generic EOM of torsion-spin}) is associated with the  spin density tensor $S^{\mu\beta\alpha}$. Equations (\ref{eq:generic EOM of metric-mass}) and (\ref{eq:generic EOM of torsion-spin}) together give the Einstein-Cartan field equations:
\begin{equation}
G^{\mu\nu} = k \;  \; \Sigma^{\mu\nu}, \ 
\label{TPS1}
\end{equation}
\begin{equation}
T^{\mu\beta\alpha} =k \;   \; \tau^{\mu\beta\alpha}.
\label{TPS2}
\end{equation}
$G^{\mu\nu}$ is the asymmetric Einstein tensor constructed from the asymmetric connection. $\Sigma^{\mu\nu}$ is the canonical energy-momentum tensor (asymmetric) constructed from the metric energy-momentum tensor (symmetric) and the spin density tensor. In Eq. (\ref{TPS2}), $T^{\mu\beta\alpha}$ is the modified torsion (traceless part of the torsion tensor); it is algebraically related to $S^{\mu\beta\alpha}$ on the right hand side. On setting the torsion to zero, the field equations of general relativity are recovered.

For a Dirac field ($\psi$), the matter lagrangian density is given by
\begin{equation}\label{diraclagrangian}
\mathcal{L}_{m} = \frac{i\hbar c}{2}(\overline{\psi}\gamma^{\mu}\nabla_{\mu}\psi - \nabla_{\mu}\overline{\psi}\gamma^{\mu}\psi) - mc^{2}\overline{\psi}\psi.
\end{equation}
We denote a Riemannian space-time by $V_4$ and a space-time with torsion by $U_4$. Minimally coupling a Dirac field on $U_4$ leads to the Einstein-Cartan-Dirac (ECD) theory. The spinors are defined on $V_4$ and $U_4$ using tetrads. We use $\hat{e}^{\mu} = {\partial^{\mu}}$ as the coordinate basis, which is covariant under general coordinate transformations. Spinors (defined on a Minkowski space-time) on the other hand  are associated with basis vectors which are covariant under local Lorentz transformations. To this aim, we define at each point on the manifold, four orthonormal basis fields (tetrad fields) ${\hat{e}^i (x)}$, one for each $i$ value. The tetrad fields satisfy the relation ${\hat{e}^i (x)} = e^i_{\mu} (x) \hat{e}^{\mu}$, where the transformation matrix $e^i_{\mu}$ is such that
\begin{equation}\label{eq:tetrad-metric transformation}
e^{(i)}_{\mu} e^{(k)}_{\nu} \eta_{{(i)}{(k)}} = g_{\mu\nu}.
\end{equation}
The trasformation matrix $e^{(i)}_{\mu}$ facilitates the conversion of the components of any world tensor (which transform according to general coordinate transformations) to the corresponding components in a local Minkowski space (these latter components being covariant under local Lorentz transformations). Greek indices are raised and lowered using the metric $g_{\mu\nu}$, while the Latin indices are raised and lowered using $\eta_{(i)(k)}$. Parentheses around indices is a matter of convention. 

We adopt the following conventions for the remainder of the paper:
\begin{itemize}
	\item Objects with Greek indices (world indices), e.g. $\alpha, \zeta, \delta $, transforms according to \textit{general coordinate transformations} and are raised and lowered using the metric $g_{\mu\nu}$.
	\item Objects with Latin indices within parentheses (tetrad indices), e.g. (a) or (i), transform according to \textit{local Lorentz transformations} and are raised and lowered using $\eta_{(i)(k)}$.
	\item Latin indices without parentheses, e.g. $i,j,b,c$ refer to objects in Minkowski space, which transform according to \textit{global Lorentz transformations}.
	\item In general $0,1,2,3$ refer to world indices while $(0),(1),(2),(3)$ refer to tetrad indices.
	\item $\nabla^{\{\}}$ represents the covariant derivative with the Christoffel connections ($\{\}$), while $\nabla$ denotes the total covariant derivative.
	\item Commas $(,)$ refer to partial derivatives and semicolons $(;)$ to the Riemannian covariant derivative, which implies ($;$) and $\nabla^{\{\}}$ are the same for tensors. For spinors, $(;)$ involves a partial derivative and the Riemannian part of the spin connection.
\end{itemize}

Just as the affine connection $\Gamma$ facilitates parallel transport of geometrical objects with world (Greek) indices, the spin connection ($\gamma$) does so for anholonomic objects (those with Latin indices). The affine connection $\Gamma$ has two parts - Riemannian ($\{\}$) and torsional (constructed from the contorsion tensor $K_{\mu}^{\:\:\:(k)(i)}$), similarly, the spin connection ($\gamma_{\mu}^{\:\:\:(i)(k)}$) has two parts - Riemannian (denoted by $\gamma_{\mu}^{o\:\:\:(i)(k)}$) and torsional (constructed from the contorsion tensor $K_{\mu}^{\:\:\:(k)(i)}$). These quantities are interrelated by
\begin{equation}\label{eq:total spin connection as sum of riemannian and torsional part}
\gamma_{\mu}^{\:\:\:(i)(k)} = \gamma_{\mu}^{o\:\:\:(i)(k)} - K_{\mu}^{\:\:\:(k)(i)} \ \text{and}
\end{equation}
\begin{equation}\label{eq:relation between spin conection and affine connection is as follows}
\begin{split}
\gamma_{\mu}^{\:\:\:(i)(k)} &= e^{(i)}_{\alpha}e^{\nu (k)} \Gamma_{\mu\nu}^{\:\:\:\:\:\:\:\alpha} - e^{\nu (k)}\partial_{\mu}e^{(i)}_{\nu}\\ &= e^{(i)}_{\alpha}e^{\nu (k)} \bfrac{\alpha}{\mu \nu} - K_{\mu}^{\:\:\:(k)(i)} - e^{\nu(k)} \partial_{\mu}e^{(i)}_{\nu}.
\end{split}
\end{equation}

Using equations (\ref{eq:total spin connection as sum of riemannian and torsional part}) and (\ref{eq:relation between spin conection and affine connection is as follows}) , the Riemannian part of the spin connection can be expressed entirely in terms of the Christoffel symbols and the tetrads as \cite{gasperini}
\begin{equation}\label{26}
\bfrac{\alpha}{\mu \nu} = e_{(i)}^{\alpha}e_{\nu (k)} \gamma_{\mu}^{o\:\:\:(k)(i)} +  e_{(i)}^{\alpha}\partial_{\mu}e^{(i)}_{\nu}.
\end{equation}
One can thus define the covariant derivative for spinors as 
\begin{align}
&\psi_{;\mu} = \partial_{\mu} \psi + \frac{1}{4}\gamma^o_{\mu (b)(c)}\gamma^{[(b)}\gamma^{(c)]}\psi \: \: \: \: \: \: \: \: \: \: \: \: \: \: \: \: \: \: \: \: \: \: \: \: \: \: \: \: \: \: \: \: \: \: \: \: \: \: \: \: \: \: \: \: \: \: \: \: \: \: \: \: \: \: \: \: (\text{on} \: \: V_4) \label{covariantderivativev4}\\
\text{and} \ &\nabla_{\mu} \psi = \partial_{\mu} \psi +  \frac{1}{4}\gamma^0_{\mu (c)(b)}\gamma^{[(b)}\gamma^{(c)]}\psi - \frac{1}{4} K_{\mu (c)(b)}\gamma^{[(b)}\gamma^{(c)]}\psi. \: \: \: \: \: \: \: \: \: \: \: (\text{on} \: \: U_4) \label{covariantderivative}
\end{align}
\indent The explicit form of the matter Lagrangian density is obtained by substituting equations (\ref{covariantderivativev4}) and (\ref{covariantderivative}) in Eq. (\ref{diraclagrangian}). The Dirac equation is then given by Eq. (\ref{eq:generic EOM of matter}):
\begin{align}
&i\gamma^{\mu}\psi_{;\mu}- \frac{mc}{\hbar}\psi = 0 \: \: \: \: \: \: \: \: \: \: \: \: \: \: \: \: \: \: \: \: \: \: \: \: \: \: \: \: \: \: \: \: \: \: \: \: \: \: \: \: \: \: \: \: \: \: \: \: \: \: \: \: \: \: \: \: \: \: \: \: \: \: \: \: \: \: \: \: \: \: \: \: \: \: \: \: \: \: \: \: \: (\text{on} \: \: V_4) \label{eq:DE V4}\\
\text{and} \ &i\gamma^{\mu}\psi_{;\mu} + \frac{i}{4}K_{(a)(b)(c)}\gamma^{[(a)}\gamma^{(b)}\gamma^{(c)]}\psi - \frac{mc}{\hbar}\psi = 0.  \: \: \: \: \: \: \: \: \: \: \: \: \: \: \: \: \: \: \: \: \: \: \: \: \: \: \: \: \: \:(\text{on} \: \: U_4)\label{eq:DE U4}
\end{align}
\indent The gravitational field equations are obtained using equations (\ref{eq:generic EOM of metric-mass})
and (\ref{diraclagrangian}):
\begin{align}
&G_{\mu\nu}(\{\}) = \frac{8\pi G}{c^4} T_{\mu\nu} \: \: \: \: \: \: \: \: \: \: \: \: \: \: \: \: \: \: \: \: \: \: \: \: \: \: \: \: \: \: \: \: \: \: \: \: \: \: \: \: \: \: \: \: \: \: \: \: \: \: \: \: \: \: \: \: \: \: \: \: \: \: \: \: \: \: \: \: \: \: \: \: \: \: \: \: \: \: \: \: (\text{on} \: \: V_4)    \label{eq:efev4}\\
\text{and} \ &G_{\mu\nu}(\{\}) = \frac{8\pi G}{c^4} T_{\mu\nu} - \frac{1}{2} \bigg{(}\frac{8\pi G}{c^4}\bigg{)}^2 g_{\mu\nu}  S^{\alpha\beta\lambda}S_{\alpha\beta\lambda}. \: \: \: \: \: \: \: \:\: \: \: \: \: \: \: \: \: \:\: \: \: \: \: \: \: \: \: \: (\text{on} \: \: U_4)   \label{eq:efeu4}
\end{align}
\indent The metric EM tensor (symmetric) is defined by
\begin{equation}\label{dynamic EM tensor}
T_{\mu\nu} = \Sigma_{(\mu\nu)}(\{\}) = \frac{i\hbar c}{4}\Big[\bar{\psi}\gamma_{\mu} \psi_{;\nu} + \bar{\psi}\gamma_{\nu} \psi_{;\mu} - \bar{\psi}_{;\mu} \gamma_{\nu}\psi -\bar{\psi}_{;\nu} \gamma_{\mu}\psi  \Big].
\end{equation}
\indent Equations (\ref{eq:DE V4}) and (\ref{eq:efev4}) are the governing equations for the Einstein-Dirac theory. The spin density tensor is obtained form the matter Lagrangian density (\ref{diraclagrangian}):
\begin{equation}\label{eq:spindensity}
S^{\mu\nu\alpha} = \frac{-i\hbar c}{4}\bar{\psi}\gamma^{[\mu}\gamma^{\nu}\gamma^{\alpha]}\psi.
\end{equation}
\indent Using equations (\ref{eq:spindensity}) and (\ref{eq:generic EOM of torsion-spin}),
Eq. (\ref{eq:DE U4}) simplifies to the Hehl-Datta equation \cite{hehl_RMP, Hehl1971} which together with Eq. (\ref{eq:efeu4}) and the relation between the modified torsion tensor and the spin density tensor, constitutes the field equations for the Einstein-Cartan-Dirac theory:
\begin{align}
G_{\mu\nu}(\{\}) &= \frac{8\pi G}{c^4} T_{\mu\nu} - \frac{1}{2} \bigg{(}\frac{8\pi G}{c^4}\bigg{)}^2 g_{\mu\nu}  S^{\alpha\beta\lambda}S_{\alpha\beta\lambda},\label{eq:ECDgravity}\\
T_{\mu\nu\alpha} &=  - K_{\mu\nu\alpha} = \frac{8\pi G}{c^4}S_{\mu\nu\alpha} \ \text{and} \label{eq:ECDtorsion}\\
i\gamma^{\mu}\psi_{;\mu} &= +\frac{3}{8}L_{Pl}^{2}\overline{\psi}\gamma^{5}\gamma_{(a)}\psi\gamma^{5}\gamma^{(a)}\psi + \frac{mc}{\hbar}\psi.   \label{eq:HD}
\end{align}  
Here, $L_{Pl}$ is the Planck length. The Lorentz signature, diag(+, -, -, -) is used througout the paper.
The gamma matrices are represented in the Dirac basis, which happens to be the matrix representation of Clifford algebra $Cl_{1,3}[\mathbb{R}]$:
\begin{align}
\gamma^0 = \beta = \begin{pmatrix}
\mathbb{I}_2 & 0 \\ 0 & -\mathbb{I}_2
\end{pmatrix}, \gamma^i = \begin{pmatrix}
0 & \sigma^i\\ -\sigma^i & 0 \end{pmatrix}, \gamma^5 = \frac{i}{4!}\epsilon_{ijkl}\gamma^i\gamma^j\gamma^k\gamma^l = \begin{pmatrix}
0 & \mathbb{I}_2 \\ \mathbb{I}_2 & 0 \end{pmatrix} \ \text{and} \ \alpha^i = \beta\gamma^i = \begin{pmatrix}
0 & \sigma^i\\ \sigma^i & 0 \end{pmatrix}.
\end{align}
\section{Non-relativistic limit of the Einstein-Dirac equations}
\subsection{Ansatz for the spinor and the metric}
{Ansatz for the Dirac spinor:} We expand $\psi (x,t)$ as  $\psi (x,t) = e^{iS(x,t)\hbar}$ (which can be done for any complex function of $x$ and $t$). $S$ can be expressed as a perturbative power series in $\sqrt{\hbar}$ or $(1/c)$,  to obtain the semi-classical and the non-relativistic limit respectively. The scheme for obtaining the non-relativistic limit has been employed by Kiefer and Singh \cite{kiefer_Singh_1991}. Giulini and Großardt in their work \cite{guilini_grosardt} construct a new ansatz with the parameter $\sqrt{\hbar}/{c}$ as follows:
\begin{equation}\label{eq:Spinor_ansatz_NRlimit}
	\psi(\textbf{r},t) = e^{\frac{ic^2}{\hbar}S(\textbf{r},t)}\sum_{n=0}^{\infty}\bigg(\frac{\sqrt{\hbar}}{c}\bigg)^n a_n(\textbf{r},t),
\end{equation}
where $S(\textbf{r},t)$ is a scalar function and $a_n(\textbf{r},t)$ is a spinor field. We use this ansatz in our present work.

{Ansatz for the metric:}
We express a general metric as a perturbative power series in the parameter ${\sqrt{\hbar}}/{c}$, similar to the expansion for the spinor:
\begin{equation} \label{eq:metric_ansatz_NRlimit_generic}
	g_{\mu\nu}(\textbf{r}, t) = \eta_{\mu\nu} + \sum_{n = 1}^{\infty} \bigg{(}\frac{\sqrt{\hbar}}{c} \bigg{)}^n g_{\mu\nu}^{[n]}(\textbf{r}, t),
\end{equation}
where $g_{\mu\nu}^{[n]}(x)$ are metric functions indexed by $n$. In the non-relativistic scheme, gravitational potentials are weak and cannot produce velocities comparable to $c$. Hence, we assume the leading order function to be the Minkowski metric,  $g_{\mu\nu}^{[0]}(x) = \eta_{\mu\nu}$. The generic power series for the tetrads, spin coefficients and Einstein tensor are then given by
\begin{align}
	e^{\mu}_{(i)} &=  \delta^{\mu}_{(i)} + \sum_{n = 1}^{\infty} \bigg{(}\frac{\sqrt{\hbar}}{c} \bigg{)}^n e^{\mu[n]}_{(i)}, ~~~~~~~~~~~~~~~~~~~~~~ \gamma_{(a)(b)(c)} = \sum_{n = 1}^{\infty} \bigg{(}\frac{\sqrt{\hbar}}{c} \bigg{)}^n \gamma^{[n]}_{(a)(b)(c)}, \label{eq:geeric expansions1}\\
	e^{(i)}_{\mu} &=\delta^{(i)}_{\mu} + \sum_{n = 1}^{\infty} \bigg{(}\frac{\sqrt{\hbar}}{c} \bigg{)}^n e_{\mu}^{(i)[n]} ~~~~~~~~~\text{and}~~~~~~~~~ G_{\mu\nu} = \sum_{n = 1}^{\infty} \bigg{(}\frac{\sqrt{\hbar}}{c} \bigg{)}^n G^{[n]}_{\mu\nu}, \label{eq:geeric expansions2}
\end{align}
where  $e^{\mu[n]}_{(i)} $, $e_{\mu}^{(i)[n]}$, $\gamma^{[n]}_{(a)(b)(c)}$ and $G^{[n]}_{\mu\nu}$ are functions of the metric $g_{\mu\nu}^{[n]}$ and its derivatives.



\subsection{Analyzing the Dirac equation with the above ansatz}\label{sec:DE with ansatz}
We first separate the spatial and the temporal part of the Dirac equation on $V_4$ (Eq. \ref{eq:DE V4}). [Note that $\gamma^{(a)}\psi_{;(a)}= e^{(a)}_{\mu} e^{\nu}_{(a)} \gamma^{\mu}\psi_{;\nu} = \delta_{\mu}^{\nu}\gamma^{\mu}\psi_{;\nu} = \gamma^{\mu}\psi_{;\mu}$].
\begin{align}\label{eq:DE}
	&i\gamma^{\mu}\psi_{;\mu} - \frac{mc}{\hbar} \psi = 0 \\
	\Rightarrow ~~ & i\gamma^{0}\partial_{0}\psi + \frac{i}{4}\gamma^{(0)}\gamma^o_{(0)(b)(c)}\gamma^{[(b)}\gamma^{(c)]}\psi + i\gamma^{\alpha}\partial_{\alpha}\psi + \frac{i}{4}\gamma^{(j)}\gamma^o_{(j) (b)(c)}\gamma^{[(b)}\gamma^{(c)]}\psi - \frac{mc}{\hbar}\psi = 0.  \label{Ieq:4.1}
\end{align}
Multiplying both sides by $e^{(0)}_{0}\gamma^{(0)}c$, we get 
\begin{align}\label{Ieq:4.2}
	i\partial_{t}\psi + \frac{ic}{4}\gamma^o_{0(b)(c)}\gamma^{[(b)}\gamma^{(c)]}\psi + ic~e^{(0)}_{0}e^{\alpha}_{(a)}\alpha^{(a)}\partial_{\alpha}\psi + \frac{ic}{4} e^{(0)}_{0}\alpha^{(j)}\gamma^o_{(j) (b)(c)}\gamma^{[(b)}\gamma^{(c)]}\psi - e^{(0)}_{0}\beta \frac{mc^2}{\hbar}\psi = 0.
\end{align}
Using the series expansion for the tetrads and the Riemannian part of the spin connection (equations (\ref{eq:geeric expansions1}) and (\ref{eq:geeric expansions2})), we keep terms of the order $c^2$, $c$ and $1$, and neglect terms of the order $\Big(\frac{1}{c^n}\Big)$ with $n\geq$1. This is sufficient for obtaining the equation obeyed by the leading order spinor term, $a_0$.  We thus obtain
\begin{align} \label{eq:DE with generic metric ansatz}
	\begin{split}
		i\partial_{t}\psi & + \frac{i\sqrt{\hbar}}{4}\gamma^{o[1]}_{0(b)(c)}\gamma^{[(b)}\gamma^{(c)]}\psi + ~ ic{\bf\alpha}.\nabla\psi + i\sqrt{\hbar}\vec{E}.\nabla\psi + \frac{i\sqrt{\hbar}}{4}\alpha^{(j)}\gamma^{o[1]}_{(j)(b)(c)}\gamma^{[(b)}\gamma^{(c)]}\psi  \\ - & \beta\frac{ mc^2}{\hbar}\psi  - \beta \frac{mc}{\sqrt{\hbar}}e^{(0)[1]}_{0}\psi - \beta m e^{(0)[2]}_{0} \psi = 0,
	\end{split}
\end{align}
where $\vec{E} = \Bigg( \Big[e^{(0)[1]}_{0} \alpha^{\mathbf{(1)}} + e^{\mathbf{1}~[1]}_{(a)}\alpha^{(a)}\Big] , \Big[e^{(0)[1]}_{0} \alpha^{(\mathbf{2})} + e^{\mathbf{2}~[1]}_{(a)}\alpha^{(a)} \Big], \Big[e^{(0)[1]}_{0} \alpha^{(\mathbf{3})} + e^{\mathbf{3}~[1]}_{(a)}\alpha^{(a)}\Big] \Bigg)$.
We now evaluate each term of Eq. (\ref{eq:DE with generic metric ansatz}) by substituting the spinor ansatz (\ref{eq:Spinor_ansatz_NRlimit}): \\
\textbf{Term 1}
\begin{align}\label{eq:term1}
	+i\partial_t\psi &= i\partial_t\Big[e^{\frac{ic^2S}{\hbar}}\sum_{n=0}^{\infty}\bigg(\frac{\sqrt{\hbar}}{c}\bigg)^n a_n\Big]\nonumber \\
	&=e^{\frac{ic^2S}{\hbar}}\frac{c^3}{\hbar^{3/2}}\sum_{n=0}^{\infty}\bigg(\frac{\sqrt{\hbar}}{c}\bigg)^n\Big[-\dot{S}a_{n-1} + i\dot{a}_{n-3}\Big].
\end{align}
\textbf{Term 2}
\begin{align}\label{eq:term2}
	+\frac{i\sqrt{\hbar}}{4}\gamma^{o[1]}_{(0)(b)(c)}\gamma^{[(b)}\gamma^{(c)]}\psi &=
	+\frac{i\sqrt{\hbar}}{4}\gamma^{o[1]}_{(0)(b)(c)}\gamma^{[(b)}\gamma^{(c)]}\Big[e^{\frac{ic^2S}{\hbar}}\sum_{n=0}^{\infty}\bigg(\frac{\sqrt{\hbar}}{c}\bigg)^n a_n\Big] \nonumber\\
	& =e^{\frac{ic^2S}{\hbar}}\frac{c^3}{\hbar^{3/2}}\sum_{n=0}^{\infty}\bigg(\frac{\sqrt{\hbar}}{c}\bigg)^n\Big[\frac{i\sqrt{\hbar}}{4} \gamma^{o[1]}_{(0)(b)(c)}\gamma^{[(b)}\gamma^{(c)]} a_{n-3}\Big].
\end{align}
\textbf{Term 3}
\begin{align}\label{eq:term3}
	ic\alpha.\nabla\psi &= ic\overrightarrow{\alpha}\cdot\overrightarrow{\nabla}\Big[e^{\frac{ic^2S}{\hbar}}\sum_{n=0}^{\infty}\bigg(\frac{\sqrt{\hbar}}{c}\bigg)^n a_n\Big]\nonumber \\
	&=ic\overrightarrow{\alpha}\cdot\Big[e^{\frac{ic^2S}{\hbar}}\frac{c^2}{\hbar}\sum_{n=0}^{\infty}\bigg(\frac{\sqrt{\hbar}}{c}\bigg)^n\Big(i\overrightarrow{\nabla}S a_n + \overrightarrow{\nabla}a_{n-2}\Big)\Big]\nonumber \\
	&=e^{\frac{ic^2S}{\hbar}}\frac{c^3}{\hbar^{3/2}}\sum_{n=0}^{\infty}\bigg(\frac{\sqrt{\hbar}}{c}\bigg)^n\Big[-\sqrt{\hbar}\overrightarrow{\alpha}\cdot\overrightarrow{\nabla}Sa_n + i\sqrt{\hbar}\overrightarrow{\alpha}\cdot\overrightarrow{\nabla}a_{n-2}\Big].
\end{align}
\textbf{Term 4}
\begin{align}\label{eq:term4}
	+i\sqrt{\hbar}\vec{E}.\vec{\nabla}\psi &=i\sqrt{\hbar}\vec{E}. \vec{\nabla} \Big[e^{\frac{ic^2S}{\hbar}}\sum_{n=0}^{\infty}\bigg(\frac{\sqrt{\hbar}}{c}\bigg)^n a_n\Big]\nonumber \\ &=e^{\frac{ic^2S}{\hbar}}\frac{c^3}{\hbar^{3/2}}\sum_{n=0}^{\infty}\bigg[ -\sqrt{\hbar}\vec{E}. \vec{\nabla}S~ a_{n-1} + i\sqrt{\hbar}\vec{E}. \vec{\nabla}a_{n-3}\bigg].
\end{align}
\textbf{Term 5}
\begin{align}\label{eq:term5}
	+\frac{i\sqrt{\hbar}}{4}\alpha^{(j)}\gamma^{o[1]}_{(j)(b)(c)}\gamma^{[(b)}\gamma^{(c)]}\psi &=
	+\frac{i\sqrt{\hbar}}{4}\alpha^{(j)}\gamma^{o[1]}_{(j)(b)(c)}\gamma^{[(b)}\gamma^{(c)]}\Big[e^{\frac{ic^2S}{\hbar}}\sum_{n=0}^{\infty}\bigg(\frac{\sqrt{\hbar}}{c}\bigg)^n a_n\Big] \nonumber\\
	& =e^{\frac{ic^2S}{\hbar}}\frac{c^3}{\hbar^{3/2}}\sum_{n=0}^{\infty}\bigg(\frac{\sqrt{\hbar}}{c}\bigg)^n\Big[\frac{i\sqrt{\hbar}}{4} \alpha^{(j)}\gamma^{o[1]}_{(j)(b)(c)}\gamma^{[(b)}\gamma^{(c)]} a_{n-3}\Big].
\end{align}
\textbf{Term 6}
\begin{align}\label{eq:term6}
	-\beta\frac{mc^2}{\hbar}\psi &= -\beta\frac{mc^2}{\hbar}e^{\frac{ic^2S}{\hbar}}\sum_{n=0}^{\infty}\bigg(\frac{\sqrt{\hbar}}{c}\bigg)^n a_n \nonumber \\
	&= -e^{\frac{ic^2S}{\hbar}}\frac{c^3}{\hbar^{3/2}}\sum_{n=0}^{\infty}\bigg(\frac{\sqrt{\hbar}}{c}\bigg)^n (\beta m a_{n-1}).
\end{align}
\textbf{Term 7}
\begin{align}\label{eq:term7}
	- \beta \frac{mc}{\sqrt{\hbar}}e^{(0)[1]}_{0}\psi &= - \beta \frac{mc}{\sqrt{\hbar}}e^{(0)[1]}_{0}\Big[e^{\frac{ic^2S}{\hbar}}\sum_{n=0}^{\infty}\bigg(\frac{\sqrt{\hbar}}{c}\bigg)^n a_n\Big] \nonumber \\  &= -e^{\frac{ic^2S}{\hbar}}\frac{c^3}{\hbar^{3/2}}\sum_{n=0}^{\infty}\bigg(\frac{\sqrt{\hbar}}{c}\bigg)^n \Big[\beta m ~ e^{(0)[1]}_{0} a_{n-2}\Big].
\end{align}
\textbf{Term 8}
\begin{align}\label{eq:term8}
	- \beta m e^{(0)[2]}_{0} \psi &=- \beta m \bigg{[}e^{(0)[2]}_{0}  \Big[e^{\frac{ic^2S}{\hbar}}\sum_{n=0}^{\infty}\bigg(\frac{\sqrt{\hbar}}{c}\bigg)^n a_n\Big] \nonumber \\  &=
	-e^{\frac{ic^2S}{\hbar}}\frac{c^3}{\hbar^{3/2}}\sum_{n=0}^{\infty}\bigg(\frac{\sqrt{\hbar}}{c}\bigg)^n \beta m e^{(0)[2]}_{0}a_{n-3}.
\end{align}
We thus obtain
\begin{equation}\label{eq:DE_generic_after substituting spinor, metric ansatz}
	\begin{split}
		e^{\frac{ic^2S}{\hbar}}\frac{c^3}{\hbar^{3/2}}\sum_{n=0}^{\infty}\bigg(\frac{\sqrt{\hbar}}{c}\bigg)^n &
		\Bigg[\Big(-\sqrt{\hbar}\vec{\alpha}.\vec{\nabla}S \Big) a_n - \Big(\dot{S} + \beta m + \sqrt{\hbar}\vec{E}.\vec{\nabla}S \Big)a_{n-1} + \Big(i\sqrt{\hbar}\vec{\alpha}.\vec{\nabla}  - \beta m ~ e^{(0)[1]}_{0} \Big)a_{n-2}\\  + \Bigg(i\partial_t + i\sqrt{\hbar}\vec{E}.\vec{\nabla} +  &\frac{i\sqrt{\hbar}}{4} \gamma^{o[1]}_{(0)(b)(c)}\gamma^{[(b)}\gamma^{(c)]} + \frac{i\sqrt{\hbar}}{4} \alpha^{(j)}\gamma^{o[1]}_{(j)(b)(c)}\gamma^{[(b)}\gamma^{(c)]} - \beta m e^{(0)[2]}_{0} \Bigg)a_{n-3} \Bigg] = 0.
	\end{split}
\end{equation}
At the leading order (n=0), we get
\begin{equation}\label{eq:nablaS=0}
	{\bf \nabla}S = 0,
\end{equation}
which implies that the scalar $S$ is a function of time only, i.e. $S = S(t)$.
The Dirac spinor is a 4-component object which can be written as $a_n = (a_{n,1},a_{n,2},a_{n,3},a_{n,4})$. We split it into  two-component spinors, $a_n^> = (a_{n,1}, a_{n,2})$ and $a_n^< = (a_{n,3}, a_{n,4})$. At order n=1, we get $\dot{S} + \beta m + \sqrt{\hbar}\vec{E}.\vec{\nabla}S = 0$. Since $\vec{\nabla}S = 0$, this implies
\begin{subequations}
	\begin{align}
		(m + \dot{S})a_0^> &= 0\label{48a}, \\
		\text{and} \ (m - \dot{S})a_0^< &= 0,\label{48b}
	\end{align}
\end{subequations}
which satisfies either $S = -mt$ with $a_0^< = 0$ or $S = +mt$ with $a_0^> = 0$. The wave function at this order is $\psi = e^{\frac{\pm imc^2t}{\hbar}} a_0$, which corresponds to particles of positive energy (lower sign) and negative energy (upper sign), at rest. We restrict ourselves to $S = -mt$ and $a_0^< = 0$, i.e. the positive energy solutions. It is implicitly assumed that the two cases (positive and negative energies) can be studied separately. We digress at this point and analyze the energy-momentum tensor.

\subsection{Analyzing the energy momentum tensor $T_{\mu\nu}$ with the above ansatz}\label{sec:T with ansatz}

The dynamical energy momentum tensor is given by Eq. (\ref{dynamic EM tensor}). We analyze all the sixteen components of $T_{\mu\nu}$:

\noindent 1) $k T_{00}$ (with the indices of the gamma matrices raised):
\begin{align} \label{k T_00_generic}
	k T_{00} = &\frac{4i\pi G\hbar}{c^4}\Bigg[\bar{\psi}\gamma^0 \Big(\partial_t\psi + \frac{c}{4}[\gamma^o_{0(i)(j)}\gamma^{[(i)}\gamma^{(j)]}]\psi \Big)- \Big(\partial_t\bar{\psi} + \frac{c}{4}[{\gamma}^o_{0(i)(j)}\gamma^{[(i)}\gamma^{(j)]}]\bar{\psi} \Big)\gamma^0\psi\Bigg]\\
	= &\frac{4i\pi G\hbar}{c^4}\Big( 1+ \sum_{n = 1}^{\infty} \big(\frac{\sqrt{\hbar}}{c} \big)^n e^{0[n]}_{(0)}\Big) \Bigg[\bar{\psi}\gamma^{(0)} \Big(\partial_t\psi + \frac{c}{4}[\gamma^o_{0(i)(j)}\gamma^{[(i)}\gamma^{(j)]}]\psi \Big)-\\ &\Big(\partial_t\bar{\psi} + \frac{c}{4}[{\gamma}^o_{0(i)(j)}\gamma^{[(i)}\gamma^{(j)]}]\bar{\psi} \Big)\gamma^{(0)}\psi\Bigg]. \nonumber
\end{align}
Substituting the spinor ansatz (Eq.\ref{eq:Spinor_ansatz_NRlimit}) in Eq.(\ref{k T_00_generic}), we obtain a series expansion for $k T_{00}$. At the leading order we get
\begin{equation}
	\begin{split}
		k T_{00}  = \frac{4i\pi G}{c^2}\bigg\{\bigg(\sum_{n=0}^{\infty}\bigg(\frac{\sqrt{\hbar}}{c}\bigg)^n a^\dagger_{n}\bigg)\bigg(\sum_{m=0}^{\infty}\bigg(\frac{\sqrt{\hbar}}{c}\bigg)^m\Big[ i\dot{S}a_m + \dot{a}_{m-2} \Big]\bigg)\\ + \bigg(\sum_{n=0}^{\infty}\bigg(\frac{\sqrt{\hbar}}{c}\bigg)^n\Big[ i\dot{S}a^{\dagger}_{n} - \dot{a}^{\dagger}_{n-2} \Big]\bigg)\bigg(\sum_{n=0}^{\infty}\bigg(\frac{\sqrt{\hbar}}{c}\bigg)^m a_{m}\bigg)  \bigg\} + \sum_{n=3}^{\infty} O\Big( \frac{1}{c^n}\Big),
	\end{split}
\end{equation}
with $(n+m = 0)$, i.e.
\begin{align}
	k T_{00}&= \frac{4\pi G i}{c^2}\bigg\{i(-m)a_{0}^{>\dagger}a_{0}^{>} + i(-m)a_{0}^{>\dagger}a_{0}^{>}\bigg\} + \sum_{n=3}^{\infty} O\Big( \frac{1}{c^n}\Big)\\
	&= \frac{8\pi G m\, |a_{0}^{>}|^2}{c^2} + \sum_{n=3}^{\infty} O\Big( \frac{1}{c^n}\Big). \label{result:T_00}
\end{align}

\noindent 2) $k T_{0\mu}$ ($\mu = 1,2,3$):
\begin{equation} \label{k T_0a_generic}
	\begin{split}
		k T_{0\mu} = \frac{2i\pi G\hbar}{c^4}&\Bigg[
		c\bar{\psi}\gamma_0\Big(\partial_{\mu}\psi + \frac{1}{4}[\gamma^o_{\mu(i)(j)}\gamma^{[(i)}\gamma^{(j)]} \psi]\Big) + c\bar{\psi}\gamma_{\mu} \Big(\partial_0\psi + \frac{1}{4}[\gamma^o_{0(i)(j)}\gamma^{[(i)}\gamma^{(j)]} \psi]\Big) \\ &-c\Big(\partial_{\mu}\bar{\psi} + \frac{1}{4}[{\gamma}^o_{\mu(i)(j)}\gamma^{[(i)}\gamma^{(j)]} \bar{\psi}]\Big)\gamma_0\psi - c\Big(\partial_0\bar{\psi} + \frac{1}{4}[{\gamma}^o_{0(i)(j)}\gamma^{[(i)}\gamma^{(j)]} \bar{\psi}]\Big)\gamma_{\mu}\psi \Bigg].
	\end{split}
\end{equation}
Terms containing the spin coefficients ($\gamma_{\mu(i)(j)}$) are of the order  $\frac{1}{c^3}$ or higher and hence do not contribute at  the order $\frac{1}{c^2}$. Rest of the terms are analyzed in appendix section \ref{sec:T_0_mu}, and are shown to have no contribution at the order $\frac{1}{c^2}$. Hence,
\begin{equation}\label{result:T_0 mu}
	k T_{0\mu} = \sum_{n=3}^{\infty} O\Big( \frac{1}{c^n}\Big).
\end{equation}
\noindent 3) $k T_{\mu\nu}$($\mu,\nu = 1,2,3$):
\begin{equation} \label{k T_mu nu_generic1}
	\begin{split}
		k T_{\mu\nu} = \frac{2i\pi G\hbar}{c^3}&\Bigg[
		+\bar{\psi}\gamma_{\mu}\Big(\partial_{\nu}\psi + \frac{1}{4}[\gamma^o_{\nu(i)(j)}\gamma^{[(i)}\gamma^{(j)]} \psi]\Big) + \bar{\psi}\gamma_{\nu} \Big(\partial_{\mu}\psi + \frac{1}{4}[\gamma^o_{\mu(i)(j)}\gamma^{[(i)}\gamma^{(j)]} \psi]\Big) \\ &-\Big(\partial_{\nu}\bar{\psi} + \frac{1}{4}[{\gamma}^o_{\nu(i)(j)}\gamma^{[(i)}\gamma^{(j)]} \bar{\psi}]\Big)\gamma_{\mu}\psi - \Big(\partial_{\mu}\bar{\psi} + \frac{1}{4}[{\gamma}^o_{\mu(i)(j)}\gamma^{[(i)}\gamma^{(j)]} \bar{\psi}]\Big)\gamma_{\nu}\psi \Bigg].
	\end{split}
\end{equation}
Once again, terms containing the spin coefficients ($\gamma_{\mu(i)(j)}$) are of the order  $\frac{1}{c^3}$ or higher and hence do not contribute at  the order $\frac{1}{c^2}$. Rest of the terms are analyzed in appendix section \ref{sec:T_mu_nu}, and are shown to have no contribution at the order $\frac{1}{c^2}$. Hence,

\begin{equation}\label{result:T_mu nu}
	k T_{\mu\nu} = \sum_{n=3}^{\infty} O\Big( \frac{1}{c^n}\Big).
\end{equation}

Results of the order analysis of the EM tensor, summarized by equations (\ref{result:T_00}) , (\ref{result:T_0 mu}) and (\ref{result:T_mu nu}) imply
\begin{align}
	\frac{|T_{00}|}{|T_{0i}|} &\ll 1, ~~~~~~ \frac{|T_{00}|}{|T_{ij}|} \ll 1,~~\text{and}~~ k|T_{00}| \sim O\Big(\frac{1}{c^2}\Big)~~~~\forall ~~ i,j \in (1,2,3).
\end{align}
Owing to Einstein's equations, the same relations then hold for the components of the Einstein tensor, i.e.
\begin{align}
	\frac{|G_{00}|}{|G_{0i}|} &\ll 1, ~~~~~~ \frac{|G_{00}|}{|G_{ij}|} \ll 1,~~\text{and}~~ |G_{00}| \sim O\Big(\frac{1}{c^2}\Big)~~~~ \forall~~ i,j \in (1,2,3). 
\end{align}

\subsection{Constraints on the metric}\label{sec:constraints on metric}

In section \ref{sec:T with ansatz} we showed that  $|G_{00}| \sim O\Big(\frac{1}{c^2}\Big)$ while all the other components of $G_{\mu\nu}$ are of higher order. For a generic metric ansatz, $G_{\mu\nu}$ has been calculated in appendix section \ref{NR:generic G}. At this point we make an important assumption -- the metric field is asymptotically flat. This leads to the following constraints on the metric components (proved in appendix section \ref{NR:constraints_asymptotoc flatness}): \\
\\
1) $G_{\mu\nu}^{[1]} = 0$ ($\forall ~ \mu,\nu$) together with the condition of asymptotic flatness of the metric, leads to the following results (proved in appendix section \ref{NR:constraints_g1}):
\begin{align}
	g_{\mu\nu}^{[1]} = 0, ~~~~~~ e^{\mu[1]}_{(i)} = 0, ~~~~~~~ e^{(i)[1]}_{\mu} = 0, ~~\text{and}~~ \gamma_{(i)(j)(k)}^{[1]} = 0 ~~~~~ \forall ~~i,j,k,\mu,\nu \in(0,1,2,3).
\end{align}
2) $G_{\mu\nu}^{[2]} = 0$ ({except for $\mu = \nu = 0$}) leads to the following constraint: $g_{\mu\nu}^{[2]} = F(\textbf{r},t) \delta_{\mu\nu}$ for some field $F(\textbf{r},t)$ (proved in appendix section \ref{NR:constraints_g2}). \\
\\
The full metric is then given by
\begin{equation}\label{eq:metric_ansatz_NRlimit_special}
	g_{\mu\nu}(\textbf{r}, t) = \begin{bmatrix} 1 & 0 & 0 & 0  \\ 0 & -1 & 0 & 0 \\ 0 & 0 &  -1& 0 \\ 0 & 0 & 0 &-1 \end{bmatrix} + \bigg{(}\frac{\hbar}{c^2}\bigg{)}\begin{bmatrix} F  & 0 & 0 & 0  \\ 0 & F  & 0 & 0 \\ 0 & 0 & F & 0 \\ 0 & 0 & 0 & F \end{bmatrix} (\textbf{r},t)+ \sum_{n = 3}^{\infty} \bigg{(}\frac{\sqrt{\hbar}}{c} \bigg{)}^n \begin{bmatrix} g^{[n]}_{00}  & g^{[n]}_{01} & g^{[n]}_{02} &g^{[n]}_{03}  \\ g^{[n]}_{10} & g^{[n]}_{11}& g^{[n]}_{12} &g^{[n]}_{13} \\ g^{[n]}_{20} &g^{[n]}_{21} & g^{[n]}_{22}&g^{[n]}_{23} \\ g^{[n]}_{30}& g^{[n]}_{31} &g^{[n]}_{32} & g^{[n]}_{33} \end{bmatrix}(\textbf{r},t),
\end{equation}
where $g^{[2]}_{00} = g^{[2]}_{11} = g^{[2]}_{22} = g^{[2]}_{33} = F(\textbf{r},t)$. The above metric has been employed to calculate other objects (tetrads, spin coefficients, etc.)
in appendix sections \ref{NR:metric}, \ref{NR:spin connection}, \ref{NR:tetrad} and \ref{NR:G}.

\subsection{Non-Relativistic (NR) limit of the Einstein-Dirac equations}\label{NR_limit_ED}
\noindent {\textbf{Dirac equation}}: In section \ref{sec:DE with ansatz} we analyzed Eq. (\ref{eq:DE_generic_after substituting spinor, metric ansatz}) for $n=0$ and $n=1$. Using the results of sections \ref{NR:metric} \ - \ \ref{NR:spin connection}, Eq. (\ref{eq:DE_generic_after substituting spinor, metric ansatz}) can be further simplified to 
\begin{align}\label{eq:DE_specific_after substituting spinor, metric ansatz}
	e^{\frac{ic^2S}{\hbar}}\frac{c^3}{\hbar^{3/2}}\sum_{n=0}^{\infty}\bigg(\frac{\sqrt{\hbar}}{c}\bigg)^n \Big[&-(\dot{S} + \beta m)~a_{n-1} + i\dot{a}_{n-3} + i\sqrt{\hbar}\overrightarrow{\alpha}\cdot\overrightarrow{\nabla}a_{n-2}  - \beta\frac{mF(\textbf{r},t)}{2}a_{n-3}\Big] = 0.
\end{align}
At order $n = 2$, Eq. (\ref{eq:DE_specific_after substituting spinor, metric ansatz}) gives us
\begin{align}\label{49}
	\left (
	\begin{tabular}{cc}
		$\dot{S}$ + m & 0 \\
		0 & $\dot{S}$ - m
	\end{tabular}
	\right )
	\left (
	\begin{array}{c}
		a_1^> \\
		a_1^<
	\end{array}
	\right )
	- i\sqrt{\hbar} \left (
	\begin{tabular}{cc}
		0 & $\overrightarrow{\sigma}\cdot\overrightarrow{\nabla}$ \\
		$\overrightarrow{\sigma}\cdot\overrightarrow{\nabla}$ & 0 
	\end{tabular}
	\right )
	\left (
	\begin{array}{c}
		a_0^> \\
		a_0^<
	\end{array}
	\right ) = 0.
\end{align}
The first of these two equations is trivially satisfied. The second equation yields a relation between $a_1^<$ and $a_0^>$:
\begin{equation}\label{50}
	a_1^< = \frac{-i\sqrt{\hbar}\overrightarrow{\sigma}\cdot\overrightarrow{\nabla}}{2m} a_0^>.
\end{equation}
At order $n = 3$, we get
\begin{align}\label{51}
	\left (
	\begin{tabular}{cc}
		$\dot{S}$ + m & 0 \\
		0 & $\dot{S}$ - m
	\end{tabular}
	\right )
	\left (
	\begin{array}{c}
		a_2^> \\
		a_2^<
	\end{array}
	\right )
	- i\sqrt{\hbar} \left (
	\begin{tabular}{cc}
		0 & $\overrightarrow{\sigma}\cdot\overrightarrow{\nabla}$ \\
		$\overrightarrow{\sigma}\cdot\overrightarrow{\nabla}$ & 0 
	\end{tabular}
	\right )
	\left (
	\begin{array}{c}
		a_1^> \\
		a_1^<
	\end{array}
	\right ) \nonumber \\
	-\left (
	\begin{tabular}{cc}
		$i\partial_t -  \frac{m F(\textbf{r},t)}{2}$ & 0 \\
		0 & $i\partial_t + \frac{m F(\textbf{r},t)}{2}$
	\end{tabular}
	\right )
	\left (
	\begin{array}{c}
		a_0^> \\
		a_0^<
	\end{array}
	\right ) = 0,
\end{align}
which comprises of two equations. Using Eq. (\ref{50}), the first of these two equations gives us
\begin{equation}\label{Shrodinger equation with F}
	i\hbar\frac{\partial a_0^>}{\partial t} = -\frac{\hbar^2}{2m} \nabla^2 a_0^> + \frac{m\hbar F(\textbf{r},t)}{2} a_0^>.
\end{equation}

\noindent{\textbf{Einstein's equations}}: The
Einstein tensor has been evaluated in appendix section \ref{NR:G}. Equating $G_{00}$ to k$T_{00}$ we get 
\begin{equation}
	\frac{\hbar \nabla^2 F(\textbf{r},t)}{c^2} + \sum_{n=3}^{\infty} O\Big( \frac{1}{c^n}\Big) = \frac{8\pi G m\, |a_{0}^{>}|^2}{c^2} + \sum_{n=3}^{\infty} O\Big( \frac{1}{c^n}\Big).
\end{equation}
At the leading order, this gives us
\begin{equation}\label{poisson equation with F}
	\nabla^2 F(\textbf{r},t) = \frac{8\pi G m\, |a_{0}^{>}|^2}{\hbar}.
\end{equation}
Recognizing the quantity $\frac{\hbar F(\textbf{r},t)}{2}$ as the Newtonian potential $\phi$, we obtain the Schr\"{o}dinger-Newton system of equations ($m\phi$ $\rightarrow$ gravitational potential energy and $m\, |a_{0}^{>}|^2$ $\rightarrow$ mass density) as the NR limit of the Einstein-Dirac equations:
\begin{align}
	i\hbar\frac{\partial a_0^>}{\partial t} &= -\frac{\hbar^2}{2m} \nabla^2 a_0^> + m \phi(\textbf{r},t) a_0^> \ \text{and} \\
	\nabla^2\phi(\textbf{r},t) &= 4\pi G m\, |a_{0}^{>}|^2 = 4\pi G \rho(\textbf{r},t)\\
	\implies i\hbar\frac{\partial a_0^>}{\partial t} &= -\frac{\hbar^2}{2m} \nabla^2 a_0^{>}  - G m^2\,\int \frac{|a_{0}^{>}({\bf r}{'}, t)|^2}{|{\bf r} - {\bf r}{'}|}d^3{\bf r}{'}a_0^>,
\end{align}
the physical picture for which has already been discussed in section \ref{intro}. This completes the derivation of the  Schr\"{o}dinger-Newton equation as the non-relativistic limit
of the Einstein-Dirac equations.

\section{Non-relativistic limit of the Einstein-Cartan-Dirac equations}\label{NR_ECD}
We shall now employ the WKB type ansatz of the previous section to the case when torsion is included. It is to be noted that the torsion of the Dirac field can be expressed directly in terms of the Dirac spinors. Once the substitution of the torsion tensor has been
done in terms of the Dirac spinors, the non-linear Dirac equation no longer makes any reference to torsion. Similarly, in the Einstein-Cartan field equations, the contribution coming from torsion can be expressed in terms of the Dirac state. Thus the 
Einstein-Cartan-Dirac system is a coupled differential system for the metric and the Dirac state - just like the Einstein-Dirac system is - only, the non-linear terms are now different. Thus the WKB ansatz used earlier can be directly used in the presence of torsion as well.

The Dirac equation on $U_4$ (also known as the Hehl-Datta equation) is given by (Eq. (\ref{eq:HD}))
\begin{equation} 
	i\gamma^{\mu}\psi_{;\mu} - \frac{3}{8}L_{Pl}^{2}\overline{\psi}\gamma^{5}\gamma_{(a)}\psi\gamma^{5}\gamma^{(a)}\psi - \frac{mc}{\hbar}\psi = 0.
\end{equation}
We have already analyzed the first and the last term on the left hand side using the ansatz for the spinor (\ref{eq:Spinor_ansatz_NRlimit}) and the metric (\ref{eq:metric_ansatz_NRlimit_special}). The second term arises due to torsion and makes the equation non-linear. We evaluate this term similar to the other two (section \ref{sec:DE with ansatz}). Multiplying the middle term by $e^{(0)}_{0}\gamma^{(0)}c$ (as was done in section \ref{sec:DE with ansatz} to obtain (\ref{Ieq:4.2}) from (\ref{Ieq:4.1})), we get
\begin{align} -e^{(0)}_{0}\gamma^{(0)}\frac{3c}{8}L_{Pl}^{2}\overline{\psi}\gamma^{5}\gamma_{(a)}\psi\gamma^{5}\gamma^{(a)}\psi& = -\frac{3c}{8} l_{Pl}^2 ~ e^{\frac{ic^2S}{\hbar}}\Big[1 + \frac{\hbar F(\textbf{r},t)}{2c^2} +\sum_{n=3}^{\infty} O\Big( \frac{1}{c^n}\Big) \Big] \nonumber \\  & \bigg(\sum_{n=0}^{\infty}\bigg(\frac{\sqrt{\hbar}}{c}\bigg)^n a^\dagger_{n}\bigg)\gamma^5 \gamma_{(a)} \bigg(\sum_{l=0}^{\infty}\bigg(\frac{\sqrt{\hbar}}{c}\bigg)^l a_l\bigg)\gamma^5 \gamma^{(a)}\bigg(\sum_{m=0}^{\infty}\bigg(\frac{\sqrt{\hbar}}{c}\bigg)^l a_m\bigg),
\end{align}
which simplifies to
\begin{equation}\label{HD NL term with all ansatz}
	e^{\frac{ic^2S}{\hbar}}\frac{c^3}{\hbar^{3/2}}\Bigg[1 + \frac{\hbar F(\textbf{r},t)}{2c^2} +\sum_{n=3}^{\infty} O\Big( \frac{1}{c^n}\Big)\Bigg]\frac{3G}{8} \bigg(\sum_{n_1,n_2,n_3=0}^{\infty}\bigg(\frac{\sqrt{\hbar}}{c}\bigg)^{n} a^{\dagger}_{n_1-i}\gamma^5\gamma_{(a)} a_{n_2-j}\gamma^5\gamma^{(a)} a_{n_3-k}\bigg),
\end{equation}
where $n = n_1 + n_2 + n_3$. This term modifies 
Eq. (\ref{eq:DE_specific_after substituting spinor, metric ansatz}) as follows:
\begin{equation}\label{eq:HD_specific_after substituting spinor, metric ansatz}
	\begin{split}
		e^{\frac{ic^2S}{\hbar}}\frac{c^3}{\hbar^{3/2}}\sum_{n=0}^{\infty}& \bigg(\frac{\sqrt{\hbar}}{c}\bigg)^n \Bigg[m~a_{n-1} + i\dot{a}_{n-3} + i\sqrt{\hbar}\overrightarrow{\alpha}\cdot\overrightarrow{\nabla}a_{n-2} - \beta m a_{n-1} - \beta\frac{mF(\textbf{r},t)}{2}a_{n-3} \\
		& + \frac{3G}{8}\bigg(\sum_{n_1,n_2,n_3=0}^{\infty}\bigg(\frac{\sqrt{\hbar}}{c}\bigg)^{n} a^{\dagger}_{n_1-i}\gamma^5\gamma_{(a)} a_{n_2-j}\gamma^5\gamma^{(a)} a_{n_3-k}\bigg) \Bigg] = 0,
	\end{split}
\end{equation}
where $n = n_1 + n_2 + n_3$ and $i+j+k = 5$, with $i\leq n_1$, $j\leq n_2$ and $k\leq n_3$. Further, $i,j,k,n_1,n_2,n_3$ $\in$ (0,1,2,3,4,5). The non-linear term contributes only at order n=5 and higher. As a result, the analysis for $n = 0,1,2$ and $3$ (considered in  section \ref{NR_limit_ED}) holds good. Thus $a_0^>$ satisfies the  Schr\"{o}dinger equation, i.e. $ i\hbar\frac{\partial a_0^>}{\partial t} = -\frac{\hbar^2}{2m} \nabla^2 a_0^> + \frac{m\hbar F(\textbf{r},t)}{2} a_0^>$.\\

Einstein's equations on $U_4$ read: $G_{\mu\nu}(\{\}) = \rchi T_{\mu\nu} - \frac{1}{2}\rchi^{2}g_{\mu\nu} S^{\alpha\beta\lambda}S_{\alpha\beta\lambda}$.
$G_{\mu\nu}(\{\})$  and $T_{\mu\nu}$ have already been analyzed in section \ref{NR_limit_ED}. The second term on the right hand side, i.e. $- \frac{1}{2}\rchi^{2}g_{\mu\nu} S^{\alpha\beta\lambda}S_{\alpha\beta\lambda}$, involves a contraction of the spin density tensor (\ref{eq:spindensity}). We consider only the first term in the series expansion of the metric, because the other terms together with the coupling constant are of orders not relevant for the NR limit. We thus obtain
\begin{align}
	\frac{-1}{2}\rchi^{2}g_{00} S^{\alpha\beta\lambda}S_{\alpha\beta\lambda} &= -g_{00}\frac{2\pi^2G^2\hbar^2}{c^6} \sum_{N=0}^{\infty}\Big(\sum_{k=0}^{\infty} \sum_{l=0}^{\infty} a_{k}^{\dagger} \gamma^{0} \gamma^{[c} \gamma ^{a} \gamma^{b]}\Big) \Big( \sum_{m=0}^{\infty} \sum_{n=0}^{\infty} a_{m}^{\dagger} \gamma^{0} \gamma_{[c} \gamma_{a} \gamma_{b]} n_m \Big) =\sum_{n=6}^{\infty} O\Big( \frac{1}{c^n}\Big),
\end{align}
which implies that this additional term does not contribute at the order $\bigg(\frac{1}{c^2}\bigg)$ on the right hand side of Eq. (\ref{eq:ECDgravity}). Hence, we once again recover Poisson's equation. Thus the Schr\"{o}dinger-Newton equation also happens to be the NR limit of the ECD theory, which implies that torsion does not contribute at the leading order. 

\section{Conclusions}
While the non-relativitic limit of the Einstein-Dirac equations for a self-gravitating Dirac field has been calculated by Giulini and Großardt \cite{guilini_grosardt}, we relax the assumption of a spherically symmetric metric in our present work. The Schr\"{o}dinger-Newton equation is obtained as the non-relativistic limit for a general metric, by considering  a perturbative series in the parameter$\Big(\frac{\sqrt{\hbar}}{c}\Big)$, for the spinor, the metric and other relevant quantities. This scheme for obtaining the non-relativistic limit follows the WKB like expansion given by Giulini and Großardt \cite{guilini_grosardt}. 

The Einstein-Cartan-Dirac equations provide an elegant system for coupling matter to the geometry of space-time, where torsion arises due to the spin of the Dirac field. The non-relativistic limit of this system of equations (derived in section \ref{NR_ECD}) yields the Schr\"{o}dinger-Newton equation, at the leading order of the parameter $\Big(\frac{1}{c}\Big)$. This suggests that torsion does not manifest itself at this order.

The effect of torsion in the higher order corrections to the Schr\"{o}dinger-Newton equation, can be obtained from the Einstein-Cartan-Dirac equations, by considering a WKB type expansion for the spinor and other relevant quantities, as was done in the present work. However, in this paper, we have restricted ourselves to the analysis at the leading order, which gives us the non-relativistic limit. A similar prescription may also be employed to obtain the higher order corrections to the Schr\"{o}dinger equation (starting from Dirac's equation) and Newton's equation for gravitation (starting from Einstein's equations).

The Einstein-Cartan-Dirac equations with the unified new length scale \cite{newlength2} provide for the possibility of a solitonic solution which interpolates between a black hole and a Dirac fermion. This is one of the primary motivations for us to study this system of field equations. The search for such solutions has been attempted in \cite{swanand2} and further work is in progress. One could well ask if Derrick's theorem \cite{Derrick} could compel such solitonic solutions to be unstable. The theorem suggests that 
stationary localized solutions to non-linear wave equations such as considered here are unstable. In the present situation however, the inclusion of torsion [which has a dispersive effect] makes it more plausible to achieve a stable balancing solution where the dispersive aspect due to torsion balances the collapse aspect due to gravity. Moreover, a way out of Derrick's no-go theorem is that the sought for solitonic solutions are periodic in time, rather than time-independent. Such solutions were actually reported by us in \cite{swanand2}. Rigorously speaking, the so-called Vakhitov-Kolokolov stability criterion \cite{vakhitov}  provides a precise condition for the linear stability of a periodic solitary wave solution. This requirement continues to hold for the Einstein-Cartan-Dirac equations as well.

\section{Acknowledgments}
\noindent
SK would like to thank the Indian Institute of Science Education and Research, Pune and the Department of Science and Technology, Govt. of India for the SHE-INSPIRE fellowship. AP and VD would like to thank the Visiting Students' Research Program 2017, organized by the Tata Institute of Fundamental Research (TIFR), Mumbai, under which a part of this work was accomplished. SK, AP and VD would like to express their sincere gratitude to the Department of Astronomy and Astrophysics, TIFR, for its hospitality. 
\section{Appendix}
\label{app:NR}
\subsection{Form of the Einstein tensor evaluated using the generic metric upto second order}\label{NR:generic G}

The ansatz for the metric is given by (Eq. (\ref{eq:metric_ansatz_NRlimit_generic}))
\begin{equation*}
	g_{\mu\nu}(x) = \eta_{\mu\nu} + \sum_{n = 1}^{\infty} \bigg{(}\frac{\sqrt{\hbar}}{c} \bigg{)}^n g_{\mu\nu}^{[n]}(x).
\end{equation*}
To the second order, the metric and its inverse is then given by

\begin{align}
	g_{\mu\nu} &= \eta_{\mu\nu}+ \Big(\frac{\sqrt{\hbar}}{c}\Big) g_{\mu\nu}^{[1]} + \Big(\frac{\hbar}{c^2}\Big) g_{\mu\nu}^{[2]} +  \sum_{n=3}^{\infty} O\Big( \frac{1}{c^n}\Big) \ \text{and} \\
	g^{\mu\nu} &= \eta^{\mu\nu} - \Big(\frac{\sqrt{\hbar}}{c}\Big) g^{\mu\nu[1]} - \Big(\frac{\hbar}{c^2}\Big)[g^{\mu[1]}_{~\beta}g^{\beta\nu[1]} + g^{\mu\nu[2]}] + \sum_{n=3}^{\infty} O\Big( \frac{1}{c^n}\Big).
\end{align}

We evaluate Christoffel symbols, Riemann curvature tensor, Ricci tensor and the scalar curvature to obtain the Einstein tensor $G_{\mu\nu}$ up to the second order as follows:

\begin{equation}
	G_{\mu\nu}(\{\}) = \Big(\frac{\sqrt{\hbar}}{c}\Big) G_{\mu\nu}^{[1]}(\{\})+ \Big(\frac{\hbar}{c^2}\Big) G_{\mu\nu}^{[2]}(\{\}),
\end{equation}
where,
\begin{align}
	G_{\mu\nu}^{[1]}(\{\}) &= -\frac{1}{2}\square \overline{g}_{\mu\nu}^{[1]}, ~~~~~ \overline{g}_{ij}^{[1]} = g_{\mu\nu}^{[1]} - \frac{1}{2}\eta_{\mu\nu} g^{[1]}, ~~~~~g^{[1]} =(\eta^{\mu\nu}g_{\mu\nu}^{[1]}), \label{eq:for G(1)}\\
	G_{\mu\nu}^{[2]}(\{\}) &= -\frac{1}{2}\square \overline{g}_{\mu\nu}^{(2)} + f(g_{\mu\nu}^{[1]}),~~~~~ \overline{g}_{ij}^{[2]} = g_{\mu\nu}^{[2]} - \frac{1}{2}\eta_{\mu\nu} g^{[2]} ~~\text{and}~~ g^{[2]} =(\eta^{\mu\nu}g_{\mu\nu}^{[2]}). \label{eq:for G(2)}
\end{align}
In Eq. (\ref{eq:for G(2)}), `$f$' is a function of $g_{\mu\nu}^{[1]}$, which is given by

\begin{align*}
	f(g^{[1]}_{\mu\nu}) = - \frac{1}{4}\Big[2\partial^{\lambda} g^{[1]}\partial_{\nu} g_{\lambda\mu}^{[1]} - 2\partial ^{\lambda} g^{[1]}\partial_{\lambda} g _{\mu\nu}^{[1]} - \partial_{\rho} g _{\nu}^{\lambda[1]} \partial_{\mu} g _{\lambda}^{\rho[1]} - \partial_{\rho} g _{\nu}^{\lambda[1]} \partial_{\lambda} g _{\mu}^{\rho[1]} +\\
	\partial_{\rho} g _{\nu}^{\lambda[1]} \partial^{\rho} g _{\lambda\mu}^{[1]} + \partial_{\nu} g_{\rho}^{\lambda[1]} \partial_{\mu} g _{\lambda}^{\rho[1]} + \partial_{\nu} g_{\rho}^{\lambda[1]} \partial_{\lambda} g _{\mu}^{\rho[1]} - \partial_{\nu} g_{\rho}^{\lambda[1]} \partial^{\rho} g_{\lambda\mu}^{[1]}\Big] \\
	-\frac{1}{8}\Big[2\partial^{\lambda} g^{[1]}\partial_{\nu} g_{\lambda\mu}^{[1]} - 2\eta_{\mu\nu}\partial^{\lambda} g^{[1]}\partial_{\lambda} g^{[1]} - \partial_{\rho} g _{\nu}^{\lambda[1]} \partial_{\mu} g _{\lambda}^{\rho[1]} - \partial_{\rho} g_{\mu}^{\lambda[1]} \partial_{\lambda} g _{\nu}^{\rho[1]} \\
	+ \partial_{\rho} g_{\mu}^{\lambda[1]} \partial^{\rho} g _{\lambda\nu}^{[1]} + \partial_{\mu} g_{\rho}^{\lambda[1]} \partial_{\nu} g_{\lambda}^{\rho[1]} + \partial_{\mu} g_{\rho}^{\lambda[1]} \partial_{\lambda} g _{\nu}^{\rho[1]} - \partial_{\nu} g_{\rho}^{\lambda[1]} \partial^{\rho} g_{\lambda\mu[1]}\Big].
\end{align*}
$G_{\mu\nu}$ happens to be the same as $G_{\mu\nu}(\{\})$ for $V_4$. For $U_4$ on the other hand, $G_{\mu\nu}(\{\})$ is the Riemannian part of  $G_{\mu\nu}$ (a symmetric tensor constructed from the Christoffel symbols). 

\subsection{Constraints on the metric due to the asymptotic flatness condition}\label{NR:constraints_asymptotoc flatness}
\subsubsection{Constraint on $g^{[1]}_{\mu\nu}$}\label{NR:constraints_g1}

From the analysis of section \ref{sec:T with ansatz} one can argue that all the components of $G^{[1]}_{\mu\nu}$ are zero, which implies $\Box \overline{g}_{\mu\nu}^{[1]} = \Box g_{\mu\nu}^{[1]} = 0$, from Eq. (\ref{eq:for G(1)}) for $\mu $ $\neq$ $\nu$ (off-diagonal terms). Gravitational waves are the non-trivial solutions to this equation.  However, they do not respect asymptotic flatness. We are therefore obliged to consider the trivial solution, i.e. $g_{\mu\nu}^{[1]} = 0$ for the off-diagonal terms. In order to evaluate the diagonal terms, we consider the following general form of the metric:

\begin{equation}
	g_{\mu\nu}^{[1]} = \begin{pmatrix}
		f_1^{[1]}&0&0&0 \\ 0&f_2^{[1]}&0&0 \\ 0&0&f_3^{[1]}&0 \\ 0&0&0&f_4^{[1]}
	\end{pmatrix}.
\end{equation}
Hence,
\begin{align}
	\bar{g}_{00}^{[1]} & = \frac{f_1^{[1]}+f_2^{[1]}+f_3^{[1]}+f_4^{[1]}}{2}, \\
	\bar{g}_{11}^{[1]} & = \frac{f_1^{[1]}+f_2^{[1]} -f_3^{[1]}- f_4^{[1]}}{2}, \\
	\bar{g}_{22}^{[1]} & = \frac{f_1^{[1]}+f_3^{[1]} -f_2^{[1]}- f_4^{[1]}}{2} \ \text{and} \\
	\bar{g}_{33}^{[1]} & = \frac{f_1^{[1]}+f_4^{[1]} -f_2^{[1]}- f_3^{[1]}}{2}. 
\end{align}
Using the above equations, we get 
\begin{align}
	\Box\bar{g}_{00}^{[1]} & = \Box\frac{f_1^{[1]}+f_2^{[1]}+f_3^{[1]}+f_4^{[1]}}{2} = 0 \Longrightarrow \Box f_1^{[1]} + \Box f_2^{[1]}+\Box f_3^{[1]} +\Box f_4^{[1]} = 0,  \label{1} \\
	\Box\bar{g}_{11}^{[1]} & = \Box\frac{f_1^{[1]}+f_2^{[1]} -f_3^{[1]}- f_4^{[1]}}{2} = 0  \Longrightarrow \Box f_1^{[1]} + \Box f_2^{[1]} = \Box f_3^{[1]} +\Box f_4^{[1]}, \label{2}\\
	\Box\bar{g}_{22}^{[1]} & = \Box\frac{f_1^{[1]}+f_3^{[1]} -f_2^{[1]}- f_4^{[1]}}{2} = 0 \Longrightarrow
	\Box f_1^{[1]} + \Box f_3^{[1]} = \Box f_2^{[1]} +\Box f_4^{[1]} \ \text{and} \label{3}\\
	\Box\bar{g}_{33}^{[1]} & = \Box\frac{f_1^{[1]}+f_4^{[1]} -f_2^{[1]}- f_3^{[1]}}{2} = 0 \Longrightarrow
	\Box f_1^{[1]} + \Box f_4^{[1]} = \Box f_2^{[1]} +\Box f_3^{[1]}. \label{4}
\end{align}
Equations (\ref{2}), (\ref{3}) and (\ref{4}) imply
\begin{align}
	\Box f_2^{[1]} &= \Box f_1^{[1]} \Longrightarrow f_2^{[1]} = f_1^{[1]} + c_1, \\
	\Box f_3^{[1]} &= \Box f_1^{[1]}  \Longrightarrow f_3^{[1]} = f_1^{[1]} + c_2 \ \text{and} \\ 
	\Box f_4^{[1]} &= \Box f_1^{[1]}  \Longrightarrow f_4^{[1]} = f_1^{[1]} + c_3 .
\end{align}
The constants $c_1, c_2$ and $c_3$ must be zero, as any one of them not being equal to zero would violate the condition of asymptotic flatness. Hence, Eq. (\ref{1}) implies, $4\Box f_1^{[1]} = 0 \Longrightarrow f_1^{[1]} = 0$ (wave solutions and non-zero constants also satisfy the equation, but do not respect asymptotic flatness). Hence, $f_i^{[1]} = 0 ~ \forall ~ i$, which in turn implies
\begin{equation}
	g_{\mu\nu}^{[1]} = 0 ~ \forall ~ \mu,\nu.
\end{equation}

\subsubsection{Constraint on $g^{[2]}_{\mu\nu}$}\label{NR:constraints_g2}
From the analysis of section \ref{sec:T with ansatz} one can argue that the off-diagonal components of $G^{[2]}_{\mu\nu}$ are zero. This implies, $\Box \overline{g}_{\mu\nu}^{[2]} = \Box g_{\mu\nu}^{[2]} = 0$ ($\mu \neq \nu$), from Eq. (\ref{eq:for G(1)}). Following the same arguments of section \ref{NR:constraints_g1}, $g_{\mu\nu}^{[2]} = 0$ ($\mu \neq \nu$) is the only allowed solution to the above equation. Once again, we consider the following general form of the metric in order to evaluate the diagonal terms:

\begin{equation}
	g_{\mu\nu}^{[2]} = \begin{pmatrix}
		f_1^{[2]}&0&0&0 \\ 0&f_2^{[2]}&0&0 \\ 0&0&f_3^{[2]}&0 \\ 0&0&0&f_4^{[2]}  
	\end{pmatrix}.
\end{equation}
Hence,
\begin{align}
	\bar{g}_{00}^{[2]} & = \frac{f_1^{[2]}+f_2^{[2]}+f_3^{[2]}+f_4^{[2]}}{2}, \\
	\bar{g}_{11}^{[2]} & = \frac{f_1^{[2]}+f_2^{[2]} -f_3^{[2]}- f_4^{[2]}}{2}, \\
	\bar{g}_{22}^{[2]} & = \frac{f_1^{[2]}+f_3^{[2]} -f_2^{[2]}- f_4^{[2]}}{2} \ \text{and} \\
	\bar{g}_{33}^{[2]} & = \frac{f_1^{[2]}+f_4^{[2]} -f_2^{[2]}- f_3^{[2]}}{2} .
\end{align}
At the second order, all the components of the Einstein tensor are zero, except for the `00' component. This implies 
\begin{align}
	\Box\bar{g}_{00}^{[2]} & = \Box\frac{f_1^{[2]}+f_2^{[2]}+f_3^{[2]}+f_4^{[2]}}{2}\Longrightarrow \Box f_1^{[2]} + \Box f_2^{[2]}+\Box f_3^{[2]} +\Box f_4^{[2]} \not= 0, \label{a} \\
	\Box\bar{g}_{11}^{[2]} & = \Box\frac{f_1^{[2]}+f_2^{[2]} -f_3^{[2]}- f_4^{[2]}}{2} \Longrightarrow \Box f_1^{[2]} + \Box f_2^{[2]} = \Box f_3^{[2]} +\Box f_4^{[2]},\label{b} \\
	\Box\bar{g}_{22}^{[2]} & = \Box\frac{f_1^{[2]}+f_3^{[2]} -f_2^{[2]}- f_4^{[2]}}{2}\Longrightarrow
	\Box f_1^{[2]} + \Box f_3^{[2]} = \Box f_2^{[2]} +\Box f_4^{[2]} \ \text{and} \label{c} \\
	\Box\bar{g}_{33}^{[2]} & = \Box\frac{f_1^{[2]}+f_4^{[2]} -f_2^{[2]}- f_3^{[2]}}{2}\Longrightarrow
	\Box f_1^{[2]} + \Box f_4^{[2]} = \Box f_2^{[2]} +\Box f_3^{[2]}.\label{d} 
\end{align}
Equations  (\ref{b}), (\ref{c}) and (\ref{d}) imply 
\begin{align}
	\Box f_2^{[2]} &= \Box f_1^{[2]} \Longrightarrow f_2^{[2]} = f_1^{[2]},\label{e} \\
	\Box f_3^{[2]} &= \Box f_1^{[2]}  \Longrightarrow f_3^{[2]} = f_1^{[2]} \ \text{and} \label{f} \\ 
	\Box f_4^{[2]} &= \Box f_1^{[2]}  \Longrightarrow f_4^{[2]} = f_1^{[2]}. \label{g}
\end{align}
The absence of constants in the above equations follows from the arguments of section \ref{NR:constraints_g1}. Using equations (\ref{e}), (\ref{f}) and  (\ref{g}), we get $f_1^{[2]} = f_2^{[2]} = f_3^{[2]} = f_4^{[2]} = F(\textbf{r},t)$, hence, 

\begin{equation}\label{h}
	g_{\mu\nu}^{[2]} = \begin{pmatrix}
		F(\textbf{r},t)&0&0&0 \\ 0&F(\textbf{r},t)&0&0 \\ 0&0&F(\textbf{r},t)&0 \\ 0&0&0&F(\textbf{r},t)
	\end{pmatrix}.
\end{equation}

\subsection{Metric and Christoffel symbols}\label{NR:metric}
\noindent The metric defined in equation (\ref{eq:metric_ansatz_NRlimit_special}) is of the form

\begin{align}
	g_{\mu\nu} &= 
	\begin{pmatrix}
		1 + \frac{\hbar F(\textbf{r},t)}{c^2} & 0 & 0 & 0 \\
		0 & -1 + \frac{\hbar F(\textbf{r},t)}{c^2} & 0 & 0 \\
		0 & 0 & -1 + \frac{\hbar F(\textbf{r},t)}{c^2} & 0 \\
		0 & 0 & 0 & -1 + \frac{\hbar F(\textbf{r},t)}{c^2}
	\end{pmatrix} +  \sum_{n=3}^{\infty} O\Big( \frac{1}{c^n}\Big) \\
	\text{and} \ g^{\mu\nu} &= 
	\begin{pmatrix}
		1 - \frac{\hbar F(\textbf{r},t)}{c^2} & 0 & 0 & 0 \\
		0 & -1 - \frac{\hbar F(\textbf{r},t)}{c^2} & 0 & 0 \\
		0 & 0 & -1 - \frac{\hbar F(\textbf{r},t)}{c^2} & 0 \\
		0 & 0 & 0 & -1 - \frac{\hbar F(\textbf{r},t)}{c^2}
	\end{pmatrix} +  \sum_{n=3}^{\infty} O\Big( \frac{1}{c^n}\Big).
\end{align}
\\
\underline{Christoffel connections}:\\
For the above metric the non-zero Christoffel connections are
\begin{align}
	\begin{aligned}\label{chris_conn}
		\Gamma^{0}_{0\mu} &= \frac{\hbar \partial_{\mu}F(\textbf{r},t)}{2c^2} +  \sum_{n=3}^{\infty} O\Big( \frac{1}{c^n}\Big),\\
		\Gamma^{\mu}_{0 0} &= \frac{\hbar \partial_{\mu}F(\textbf{r},t)}{2c^2} + \sum_{n=3}^{\infty} O\Big( \frac{1}{c^n}\Big) \ \text{and}\\
		\Gamma^{\mu}_{\mu \mu} &= \frac{-\hbar \partial_{\mu}F(\textbf{r},t)}{2c^2} +  \sum_{n=3}^{\infty} O\Big( \frac{1}{c^n}\Big),\\
	\end{aligned}
\end{align}
where $\mu$ runs from 1 to 3 (spatial coordinates only). It is worth noting that the zeroth and first order terms in $\bigg(\frac{1}{c}\bigg)$ are absent in Eq.(\ref{chris_conn}). The other non zero Christoffel connections are of order three and higher in $\bigg(\frac{1}{c}\bigg)$, which we do not mention here.

\subsection{Tetrads}\label{NR:tetrad}
Tetrads were introduced in section (\ref{EC_dirac_coupling}). For the metric defined by
\begin{align}
	dS^2 = \Bigg[1 + \frac{\hbar  F(\textbf{r},t)}{c^2}\Bigg]c^2dt^2 - \Bigg[1 - \frac{\hbar  F(\textbf{r},t)}{c^2}\Bigg] d\textbf{r}^2,
\end{align}
the tetrad fields over the entire manifold are given by
\begin{align}
	\hat{e}_{(0)} =  \frac{1}{c}\Bigg(1 + \frac{\hbar F}{c^2}\Bigg)^{\frac{1}{2}}\partial_t, ~~~
	\hat{e}_{(1)} =  \Bigg(1 - \frac{\hbar F}{c^2}\Bigg)^{\frac{1}{2}}\partial_x, ~~~
	\hat{e}_{(2)} =  \Bigg(1 - \frac{\hbar F}{c^2}\Bigg)^{\frac{1}{2}}\partial_y ~~ \text{and} ~~
	\hat{e}_{(3)} =  \Bigg(1 - \frac{\hbar F}{c^2}\Bigg)^{\frac{1}{2}}\partial_z.
\end{align}
The transformation matrices (defined in Eq. (\ref{eq:tetrad-metric transformation})) which relates the world components with the anholonomic components are given by

\begin{align}
	e_{\mu}^{(i)} &= 
	\begin{pmatrix}
		1 + \frac{\hbar F(\textbf{r},t)}{2c^2} & 0 & 0 & 0 \\
		0 & 1 - \frac{\hbar F(\textbf{r},t)}{2c^2} & 0 & 0 \\
		0 & 0 & 1 - \frac{\hbar F(\textbf{r},t)}{2c^2} & 0 \\
		0 & 0 & 0 & 1 - \frac{\hbar F(\textbf{r},t)}{2c^2} 
	\end{pmatrix}+ \sum_{n=3}^{\infty} O\Big( \frac{1}{c^n}\Big), \\
	e^{\mu}_{(i)} &= 
	\begin{pmatrix}
		1 - \frac{\hbar F(\textbf{r},t)}{2c^2} & 0 & 0 & 0 \\
		0 & 1 + \frac{\hbar F(\textbf{r},t)}{2c^2} & 0 & 0 \\
		0 & 0 & 1 + \frac{\hbar F(\textbf{r},t)}{2c^2} & 0 \\
		0 & 0 & 0 & 1 + \frac{\hbar F(\textbf{r},t)}{2c^2}   
	\end{pmatrix}+ \sum_{n=3}^{\infty} O\Big( \frac{1}{c^n}\Big), \\
	e_{\nu(k)} &= 
	\begin{pmatrix}
		1 + \frac{\hbar F(\textbf{r},t)}{2c^2} & 0 & 0 & 0 \\
		0 & -1 + \frac{\hbar F(\textbf{r},t)}{2c^2} & 0 & 0 \\
		0 & 0 & -1 + \frac{\hbar F(\textbf{r},t)}{2c^2} & 0 \\
		0 & 0 & 0 & -1 + \frac{\hbar F(\textbf{r},t)}{2c^2} 
	\end{pmatrix}+ \sum_{n=3}^{\infty} O\Big( \frac{1}{c^n}\Big) \ \text{and} \\
	e^{\nu(k)} &= 
	\begin{pmatrix}
		1 - \frac{\hbar F(\textbf{r},t)}{2c^2} & 0 & 0 & 0 \\
		0 & -1 - \frac{\hbar F(\textbf{r},t)}{2c^2} & 0 & 0 \\
		0 & 0 & -1 - \frac{\hbar F(\textbf{r},t)}{2c^2} & 0 \\
		0 & 0 & 0 & -1 - \frac{\hbar F(\textbf{r},t)}{2c^2}   
	\end{pmatrix}+ \sum_{n=3}^{\infty} O\Big( \frac{1}{c^n}\Big).
\end{align}

\subsection{Riemannian part of the spin connections ($\gamma^o_{(a)(b)(c)}$)}\label{NR:spin connection}

Using the relation between Christoffel connections and tetrad transformation matrices (Eq.(\ref{26})), the Riemannian part of the spin connections (defined by Eq.(\ref{eq:total spin connection as sum of riemannian and torsional part})) are obtained as follows:

\begin{align}
	\begin{aligned}
		\gamma^o_{(0)(0)(0)} &= \frac{-\hbar\partial_0 F}{2c^2} \frac{\Big(1+\frac{\hbar F}{ 2c^2}\Big)}{\Big(1-\frac{\hbar F}{ 2c^2}\Big)}+ \sum_{n=3}^{\infty} O\Big( \frac{1}{c^n}\Big), ~~~~~~~~~
		\gamma^o_{(i)(0)(0)} = \Big(\frac{-\hbar\partial_i F}{2c^2}\Big) \frac{\hbar F/ 2c^2}{\Big(1+\frac{\hbar F}{ 2c^2}\Big)}+ \sum_{n=5}^{\infty} O\Big( \frac{1}{c^n}\Big),\\
		\gamma^o_{(0)(i)(0)} &= \frac{-\hbar\partial_i F}{2c^2} \frac{\Big(1+\frac{\hbar F}{ 2c^2}\Big)}{\Big(1-\frac{\hbar F}{ 2c^2}\Big)}+ \sum_{n=3}^{\infty} O\Big( \frac{1}{c^n}\Big), ~~~~~~~~~
		\gamma^o_{(0)(0)(i)} =  \frac{\hbar\partial_i F}{2c^2} \frac{1}{\Big(1+\frac{\hbar F}{ 2c^2}\Big)}, \\
		\gamma^o_{(i)(i)(i)} &= \frac{\hbar\partial_i F}{2c^2} \frac{\hbar F/ 2c^2}{\Big(1+\frac{\hbar F}{ 2c^2}\Big)}+ \sum_{n=5}^{\infty} O\Big( \frac{1}{c^n}\Big), ~~~~~~~~~~~~
		\gamma^o_{(i)(i)(0)} = \gamma^o_{(i)(0)(i)} = + \sum_{n=3}^{\infty} O\Big( \frac{1}{c^n}\Big),\\
		\gamma^o_{(0)(i)(i)} &= \frac{-\hbar\partial_0F }{2c^2}+ \sum_{n=3}^{\infty} O\Big( \frac{1}{c^n}\Big), ~~~~~~~~~~~~~~~~~~~~~~
		\gamma^o_{(0)(i)(j)} = \gamma^o_{i0j} = \gamma^o_{ij0} = + \sum_{n=3}^{\infty} O\Big( \frac{1}{c^n}\Big), \\
		\gamma^o_{(i)(j)(j)} &=  \frac{-\hbar\partial_0 F}{2c^2} \frac{\Big(1-\frac{\hbar F}{ 2c^2}\Big)}{\Big(1+\frac{\hbar F}{ 2c^2}\Big)}+ \sum_{n=3}^{\infty} O\Big( \frac{1}{c^n}\Big)~~~\text{and}~~~ \gamma^o_{(i)(j)(k)} = \gamma^o_{(i)(j)(i)} = \gamma^o_{(j)(j)(i)} = + \sum_{n=3}^{\infty} O\Big( \frac{1}{c^n}\Big).
	\end{aligned}
\end{align}

The torsional part of the spin connections (defined by Eq.(\ref{eq:total spin connection as sum of riemannian and torsional part})) manifests itself as a non-linear term in the Hehl-Datta equation. This term being completely expressible in terms of the Dirac spinor, is evaluated using the spinor ansatz while deriving the non-relativistic limit of the ECD system of equations. 

\subsection{Einstein tensor}\label{NR:G}

In this section, we aim to evaluate the Einstein tensor. $G_{\mu\nu}^{[1]}$ has already been shown to be zero. Since $g_{\mu\nu}^{[1]}$ is zero, f[$g_{\mu\nu}^{[1]}$] (defined in Eq. (\ref{eq:for G(2)})) is also zero. Using $g_{\mu\nu}^{[2]}$ (defined in section (\ref{NR:constraints_g2})), $G_{\mu\nu}^{[2]}$ (Eq.\ref{eq:for G(2)}) is evaluated as follows:
\begin{equation}
	G_{\mu\nu}^{[2]} = -\frac{1}{2}\square \overline{g}_{\mu\nu}^{[2]}~~~ \text{where} ~~~\overline{g}_{\mu\nu}^{[2]} = g_{\mu\nu}^{[2]} - \frac{1}{2}\eta_{\mu\nu} (\eta^{\alpha\beta}h_{\alpha\beta}),~\text{now}
\end{equation}

\begin{equation}
	\eta^{\mu\nu}h_{\mu\nu}= \left(
	\begin{tabular}{cccc}
		1 & 0 & 0 & 0 \\
		0 & -1 & 0 & 0 \\
		0 & 0 & -1 & 0 \\
		0 & 0 & 0 & -1 
	\end{tabular}
	\right) \left(
	\begin{tabular}{cccc}
		$\frac{\hbar F(\textbf{r},t)}{c^2}$ & 0 & 0 & 0 \\
		0 &$\frac{\hbar F(\textbf{r},t)}{c^2}$ & 0 & 0 \\
		0 & 0 & $\frac{\hbar F(\textbf{r},t)}{c^2}$ & 0 \\
		0 & 0 & 0 & $\frac{\hbar F(\textbf{r},t)}{c^2}$ 
	\end{tabular}
	\right) = \frac{-2\hbar F(\textbf{r},t)}{c^2},
\end{equation}
thus $G_{\mu\nu} = 0$ for $\mu\neq \nu$. The diagonal components are given by

\begin{align}\label{A5}
	G_{00} &= -\frac{1}{2}\square \overline{g}_{00}^{[2]}
	= -\frac{\hbar}{c^2} \square F(\textbf{r},t) 
	= \Bigg[ -\frac{\hbar\partial_{t}^{2}F(\textbf{r},t)}{c^4} + \frac{\hbar\nabla^2 F(\textbf{r},t)}{c^2}\Bigg] \\
	\text{and} \ G_{\alpha\alpha} &= 0 ~~\text{because} ~~ \overline{g}_{\alpha\alpha}^{[2]} = 0 ~~\text{for}~~\alpha \in(1,2,3). 
\end{align}
Thus,
\begin{align}
	G_{\mu\nu} = \frac{\hbar}{c^2}\begin{pmatrix}
		\nabla^2 F(\textbf{r},t) & 0 & 0 & 0 \\ 0 & 0 & 0 & 0  \\ 0 & 0 & 0 & 0 \\ 0 & 0 & 0 & 0 
	\end{pmatrix} + \sum_{n=3}^{\infty} O\Big( \frac{1}{c^n}\Big).
\end{align}

\subsection{Analysis of the components of the metric EM tensor}

This section contains the calculations and proofs for some of the results used in section \ref{sec:T with ansatz}.

\subsubsection{Analysis of $kT_{0\mu}$}\label{sec:T_0_mu}

After excluding the terms containing the spin coefficients $\gamma_{\mu(i)(j)}$, $kT_{0\mu}$ (Eq. (\ref{k T_0a_generic})) is given by

\begin{align} \label{k T_0mu_generic}
	k T_{0\mu} &= \frac{2i\pi G\hbar}{c^4}\Bigg[
	c\bar{\psi}\gamma^0\partial_{\mu}\psi - c\bar{\psi}\gamma^{\mu} \partial_0\psi -c\partial_{\mu}\bar{\psi}\gamma^0\psi + c\partial_0\bar{\psi}\gamma^{\mu}\psi \Bigg] \\
	&= \frac{-2i\pi G\hbar}{c^3}\Big( 1+ \sum_{n = 1}^{\infty} \bigg{(}\frac{\sqrt{\hbar}}{c} \bigg{)}^n e^{0[n]}_{(0)}\Big) \Bigg[\bar{\psi}\gamma^{(0)}\partial_{\mu}\psi  -\partial_{\mu}\bar{\psi}\gamma^{(0)}\psi \Bigg]\label{terminator}  \\ 
	& + \frac{2i\pi G\hbar}{c^4}\Big( 1+ \sum_{n = 1}^{\infty} \bigg{(}\frac{\sqrt{\hbar}}{c} \bigg{)}^n e^{\mu[n]}_{(a)}\Big)\Bigg[ \partial_t\bar{\psi}\gamma^{(a)}\psi -\bar{\psi}\gamma^{(a)} \partial_t\psi \Bigg]. \nonumber
\end{align}
The first term on the right hand side of Eq. (\ref{terminator}) is
\begin{align}
	&= \frac{2i\pi G\hbar}{c^3}\sum_{n = 0}^{\infty} \bigg{(}\frac{\sqrt{\hbar}}{c} \bigg{)}^n\Big( a_{n_1}^{\dagger} \partial_{\mu}a_{n_2} - \partial_{\mu}a_{n_1}^{\dagger} a_{n_2}\Big) ~~~ ~~ (n=n_1+n_2)\nonumber\\
	&= \sum_{n=3}^{\infty} O\Big( \frac{1}{c^n}\Big).
\end{align}
While the second term is
\begin{align}
	\begin{split}
		&= \frac{2i\pi G}{c^2}\bigg\{\bigg(\sum_{n=0}^{\infty}\bigg(\frac{\sqrt{\hbar}}{c}\bigg)^n a^\dagger_{n}\bigg)\alpha^{(a)}\bigg(\sum_{m=0}^{\infty}\bigg(\frac{\sqrt{\hbar}}{c}\bigg)^m\Big[ i\dot{S}a_m + \dot{a}_{m-2} \Big]\bigg)\\ &+ \bigg(\sum_{n=0}^{\infty}\bigg(\frac{\sqrt{\hbar}}{c}\bigg)^n\Big[ i\dot{S}a^{\dagger}_{n} - \dot{a}^{\dagger}_{n-2} \Big]\bigg)\alpha^{(a)}\bigg(\sum_{n=0}^{\infty}\bigg(\frac{\sqrt{\hbar}}{c}\bigg)^m a_{m}\bigg)  \bigg\} + \sum_{n=3}^{\infty} O\Big( \frac{1}{c^n}\Big)
	\end{split} \nonumber \\
	&= \frac{4\pi Gm}{c^2}(a_0^{\dagger}\alpha^{(a)}a_0) + \sum_{n=3}^{\infty} O\Big( \frac{1}{c^n}\Big) \nonumber\\
	&= \frac{4\pi Gm}{c^2}\Bigg[ \begin{pmatrix} a_0^> & 0 \end{pmatrix}^{\dagger} \begin{pmatrix}0&\sigma^{(a)} \\\sigma^{(a)} &0   \end{pmatrix}\begin{pmatrix} a_0^> \\ 0 \end{pmatrix} \bigg]   + \sum_{n=3}^{\infty} O\Big( \frac{1}{c^n}\Big) \nonumber\\
	& =\sum_{n=3}^{\infty} O\Big( \frac{1}{c^n}\Big). 
\end{align}
Hence, there is no contribution at the second order.

\subsubsection{Analysis of $kT_{\mu\nu}$}\label{sec:T_mu_nu}

After excluding the terms containing the spin coefficients $\gamma_{\mu(i)(j)}$, $kT_{\mu\nu}$ (Eq. (\ref{k T_mu nu_generic1})) is given by 

\begin{align} 
	k T_{\mu\nu} = &\frac{2i\pi G\hbar}{c^3}\Bigg[
	-\bar{\psi}\gamma^{\mu}\partial_{\nu}\psi - \bar{\psi}\gamma^{\nu} \partial_{\mu}\psi +\partial_{\nu}\bar{\psi}\gamma^{\mu}\psi + \partial_{\mu}\bar{\psi}\gamma^{\nu}\psi \Bigg] \nonumber\\
	= &\frac{2i\pi G\hbar}{c^3}\Big( 1+ \sum_{n = 1}^{\infty} \bigg{(}\frac{\sqrt{\hbar}}{c} \bigg{)}^n e^{\mu[n]}_{(a)}\Big) \Bigg[\psi^{\dagger}\alpha^{(a)}\partial_{\nu}\psi  -\partial_{\nu}\psi^{\dagger}\alpha^{(a)}\psi \Bigg] \nonumber \\ 
	& + \frac{2i\pi G\hbar}{c^3}\Big( 1+ \sum_{n = 1}^{\infty} \bigg{(}\frac{\sqrt{\hbar}}{c} \bigg{)}^n e^{\nu[n]}_{(b)}\Big)\Bigg[ \partial_{\mu}\psi^{\dagger}\alpha^{(b)}\psi -\psi^{\dagger}\alpha^{(b)} \partial_{\mu}\psi \Bigg] \nonumber \\
	= &\frac{2i\pi G\hbar}{c^3}\sum_{n = 0}^{\infty} \bigg{(}\frac{\sqrt{\hbar}}{c} \bigg{)}^n\Big( e^{\mu}_{(a)}a_{n_1}^{\dagger}\alpha^{(a)} \partial_{\nu}a_{n_2} - e^{\mu}_{(a)}\partial_{\nu}a_{n_1}^{\dagger}\alpha^{(a)} + e^{\nu}_{(b)}a_{n_1}^{\dagger}\alpha^{(b)} \partial_{\mu}a_{n_2} - e^{\nu}_{(b)}\partial_{\mu}a_{n_1}^{\dagger}\alpha^{(b)} a_{n_2}\Big) \nonumber\\
	= &\sum_{n=3}^{\infty} O\Big( \frac{1}{c^n}\Big).
\end{align}
Hence, there is no contribution at the second order.

\subsection{Generic components of $T_{\mu\nu}$}\label{NR:T}

Using the spin connections of section (\ref{NR:spin connection}), we analyze the metric energy-momentum tensor (Eq.(\ref{dynamic EM tensor})), whose components are given on the following page.

\def\baselinestretch{1}\selectfont
\begin{landscape}
	\begin{equation}\label{62}
		T_{\mu\nu} = \frac{i\hbar c}{4}\left(
		\centering
		\resizebox{1.4\textwidth}{!}{%
			\begin{tabular}{cccc}
				$\begin{split}2\bar{\psi}\gamma_0(\partial_0\psi\\ + \frac{1}{4}[\gamma_{00\alpha}\gamma^0\gamma^{\alpha} + \gamma_{0\alpha 0}\gamma^{\alpha}\gamma^0]\psi)\\-(\partial_0\bar{\psi} + \frac{1}{4}[\gamma_{00\alpha}\gamma^0\gamma^{\alpha}\\ + \gamma_{0\alpha 0}\gamma^{\alpha}\gamma^0]\bar{\psi})2\gamma_0\psi\end{split}$  & $\begin{split}\bar{\psi}\gamma_0\partial_1\psi + \bar{\psi}\gamma_1(\partial_0\psi\\ + \frac{1}{4}[\gamma_{00\alpha}\gamma^0\gamma^{\alpha} + \gamma_{0\alpha 0}\gamma^{\alpha}\gamma^0]\psi)\\ - \partial_1\bar{\psi}\gamma_0\psi - (\partial_0\bar{\psi} + \frac{1}{4}[\gamma_{00\alpha}\gamma^0\gamma^{\alpha}\\ + \gamma_{0\alpha 0}\gamma^{\alpha}\gamma^0]\bar{\psi})\gamma_1\psi ) \end{split}$ & $\begin{split}\bar{\psi}\gamma_0\partial_2\psi + \bar{\psi}\gamma_2(\partial_0\psi\\ + \frac{1}{4}[\gamma_{00\alpha}\gamma^0\gamma^{\alpha} + \gamma_{0\alpha 0}\gamma^{\alpha}\gamma^0]\psi)\\ - \partial_2\bar{\psi}\gamma_0\psi - (\partial_0\bar{\psi} + \frac{1}{4}[\gamma_{00\alpha}\gamma^0\gamma^{\alpha}\\ + \gamma_{0\alpha 0}\gamma^{\alpha}\gamma^0]\bar{\psi})\gamma_2\psi ) \end{split}$ & $\begin{split}\bar{\psi}\gamma_0\partial_3\psi + \bar{\psi}\gamma_3(\partial_0\psi\\ + \frac{1}{4}[\gamma_{00\alpha}\gamma^0\gamma^{\alpha} + \gamma_{0\alpha 0}\gamma^{\alpha}\gamma^0]\psi)\\ - \partial_3\bar{\psi}\gamma_0\psi - (\partial_0\bar{\psi} + \frac{1}{4}[\gamma_{00\alpha}\gamma^0\gamma^{\alpha}\\ + \gamma_{0\alpha 0}\gamma^{\alpha}\gamma^0]\bar{\psi})\gamma_3\psi ) \end{split}$ \\ \\
				$\begin{split}\bar{\psi}\gamma_1(\partial_0\psi\\ + \frac{1}{4}[\gamma_{00\alpha}\gamma^0\gamma^{\alpha} + \gamma_{0\alpha 0}\gamma^{\alpha}\gamma^0]\psi)\\
				+ \bar{\psi}\gamma_0\partial_1\psi - (\partial_0\bar{\psi} + \frac{1}{4}[\gamma_{00\alpha}\gamma^{0}\gamma^{i}\\ + \gamma_{0\alpha 0}\gamma^{i}\gamma^{0}]\bar{\psi})\gamma_1\psi - \partial_1\bar{\psi}\gamma_0\psi\end{split}$ & $2(\bar{\psi}\gamma_1\partial_1\psi - \partial_1\bar{\psi\gamma_1\psi})$ & $\begin{split}\bar{\psi}\gamma_1\partial_2\psi + \bar{\psi}\gamma_2\partial_1\psi\\ - \partial_2\bar{\psi}\gamma_1\psi - \partial_1\bar{\psi}\gamma_2\psi\end{split}$ & $\begin{split}\bar{\psi}\gamma_1\partial_3\psi + \bar{\psi}\gamma_3\partial_1\psi\\ - \partial_3\bar{\psi}\gamma_1\psi - \partial_1\bar{\psi}\gamma_3\psi\end{split}$ \\ \\
				$\begin{split}\bar{\psi}\gamma_2(\partial_0\psi\\ + \frac{1}{4}[\gamma_{00\alpha}\gamma^0\gamma^{\alpha} + \gamma_{0\alpha 0}\gamma^{\alpha}\gamma^0]\psi)\\ + \bar{\psi}\gamma_0\partial_2\psi - (\partial_0\bar{\psi} + \frac{1}{4}[\gamma_{00\alpha}\gamma^{0}\gamma^{i}\\ + \gamma_{0\alpha 0}\gamma^{i}\gamma^{0}]\bar{\psi})\gamma_2\psi - \partial_2\bar{\psi}\gamma_0\psi\end{split}$ & $\begin{split}\bar{\psi}\gamma_2\partial_1\psi + \bar{\psi}\gamma_1\partial_2\psi\\ - \partial_1\bar{\psi}\gamma_2\psi - \partial_2\bar{\psi}\gamma_1\psi\end{split}$ & $2(\bar{\psi}\gamma_2\partial_2\psi - \partial_2\bar{\psi\gamma_2\psi})$ &  $\begin{split}\bar{\psi}\gamma_2\partial_3\psi + \bar{\psi}\gamma_3\partial_2\psi\\ - \partial_3\bar{\psi}\gamma_2\psi - \partial_2\bar{\psi}\gamma_3\psi\end{split}$ \\ \\
				$\begin{split}\bar{\psi}\gamma_3(\partial_0\psi\\ + \frac{1}{4}[\gamma_{00\alpha}\gamma^0\gamma^{\alpha} + \gamma_{0\alpha 0}\gamma^{\alpha}\gamma^0]\psi)\\ + \bar{\psi}\gamma_0\partial_3\psi - (\partial_0\bar{\psi} + \frac{1}{4}[\gamma_{00\alpha}\gamma^{0}\gamma^{i}\\ + \gamma_{0\alpha 0}\gamma^{i}\gamma^{0}]\bar{\psi})\gamma_3\psi - \partial_3\bar{\psi}\gamma_0\psi\end{split} $ & $\begin{split}\bar{\psi}\gamma_3\partial_1\psi + \bar{\psi}\gamma_1\partial_3\psi\\ - \partial_1\bar{\psi}\gamma_3\psi - \partial_3\bar{\psi}\gamma_1\psi\end{split} $ & $\begin{split}\bar{\psi}\gamma_3\partial_2\psi + \bar{\psi}\gamma_2\partial_3\psi\\ - \partial_2\bar{\psi}\gamma_3\psi - \partial_3 \bar{\psi} \gamma_2\psi\end{split}$ & $2(\bar{\psi}\gamma_3\partial_3\psi - \partial_3\bar{\psi\gamma_3\psi})$ \\ \\
			\end{tabular}%
		} 
		\right)
	\end{equation}\\
\end{landscape}

\end{document}